\documentclass[12pt,letterpaper]{JHEP3}
\usepackage{cite}
\usepackage{epsfig}
\usepackage{amsmath}
\usepackage{amssymb}

\title{The Entropic Landscape}

\author{Raphael Bousso$^{a,b,c}$ and Roni Harnik$^d$\\ \\
  $^a$ Center for Theoretical Physics, Department of Physics\\
\ \  University of California, Berkeley, CA 94720-7300, U.S.A.\\
$^b$  Lawrence Berkeley National Laboratory, Berkeley, CA 94720-8162,
  U.S.A.\\
$^c$  Institute for the Physics and Mathematics of the Universe,\\ 
\ \ University of Tokyo,
5-1-5 Kashiwa-no-Ha, Kashiwa City, Chiba 277-8568, Japan  \\   
 $^d$ Stanford Institute for Theoratical Physics,\\ 
\ \ Department of Physics, Stanford University, CA 94305, U.S.A.}

\abstract{We initiate a quantitative exploration of the entire landscape.  Predictions thus far have focused on subsets of landscape vacua that share most properties with our own.  Using the entropic principle (the assumption that entropy production traces the formation of complex structures such as observers), we derive six predictions that apply to the whole landscape.  Typical observers find themselves in a flat universe, at the onset of vacuum domination, surrounded by a recently produced bath of relativistic quanta.  These quanta are neither very dilute nor condensed, and thus appear as a roughly thermal background.  Their characteristic wavelength is of order the inverse fourth root of the vacuum energy.   These predictions hold for completely arbitrary observers, in arbitrary vacua with potentially exotic particle physics and cosmology.  They agree with observation:  We live in a flat universe at the onset of vacuum domination, whose dominant entropy production process (the glow of galactic dust) has recently produced a radiation bath (the cosmic infrared background).  This radiation is marginally dilute, relativistic, and has a wavelength of order 100 microns, as predicted.}

\begin{document}

\section{Introduction}
\label{sec-intro}

\subsection{Predictions in the multiverse}

In a theory that gives rise to an eternally inflating multiverse, such as string theory~\cite{BP,KKLT}, we must compute not what is typical, but what is typically observed. Cosmic horizons prevent observers from seeing all but a finite portion of the multiverse, near the vantage point at which they happen to find themselves.  The likely properties of their local environment depend not only on the relative abundance of different properties in the landscape of vacua, but also on correlations between these properties (such as the amount of vacuum energy) and the local abundance of observers.

It is impractical to compute the number of observers in different regions from first principles.  However, one may argue that the number of observers of a certain class is proportional to the abundance of some object necessary for their existence, such as galaxies, stars, carbon nuclei, complex organic molecules, etc.  (This class of observers had better include us, for otherwise our observations would not test the computed predictions.) This reasoning excludes from consideration observers that are outside the specific class, and by extension, all vacua in which only such observers arise.  Nevertheless, it is perfectly sensible. By computing what it typically observed by observers that are, in some way, like us, we may well falsify a given multiverse theory.   If our observations turn out to be atypical even within a restricted class of observers and vacua (for example, if we found that the observed value of the cosmological constant is many standard deviations from the mean value found by observers that live on galaxies), then the theory will be disfavored at the corresponding level of confidence.

Conditioning on observers ``like us'' does impose limitations.   What we condition on, we cannot predict.  When we restrict attention to a special class of observers and to a small fraction of the landscape, we cannot explain why we find ourselves in this particular class.  And once a multiverse theory has been supported by showing that our observations are typical in some restricted class, one would like to test the theory further by asking whether they are typical in a wider class of observers.  The fewer conditions we impose, the more is left to predict, and the more opportunities we will have to falsify the theory.  

Using the causal diamond\footnote{The causal diamond is the intersection of the causal past of a maximally extended timelike geodesic (in a long-lived de~Sitter vacuum, this is simply the interior of the event horizon) with the future of a point on the reheating surface.  The causal diamond measure is reviewed in Appendix~\ref{sec-measure}.} measure~\cite{Bou06} for regulating the divergences of eternal inflation, a prediction that transcends this limitation was recently derived~\cite{BouHar07}: Observers that live at a time $t_{\rm obs}$ after the formation of their vacuum bubble are likely to measure a cosmological constant of order
\begin{equation}
\Lambda\sim t_{\rm obs} ^{-2}~.
\label{eq-ltobs}
\end{equation}
With $t_{\rm obs}\sim 10^{10}\,\mbox{yr}\sim 10^{61}$, the time at which we happen to live, the predicted value of $\Lambda$ agrees well with the observed value,
\begin{equation}
\Lambda_o=3.7\times 10^{-122}~.
\end{equation}
(Unless explicit units are displayed, Planck units are used in this paper.)

Though this point was not emphasized in Ref.~\cite{BouHar07}, we note here that the relation (\ref{eq-ltobs}) is (statistically) predicted for {\em all\/} observers, whether or not they are in any sense ``like us'', and whether or not they live in vacua where words like ``galaxy'' or ``carbon'' even make sense.  In particular, Eq.~(\ref{eq-ltobs}) is predicted to hold independently of the actual value of $t_{\rm obs}$.  Thus, the result solves the coincidence (``why now'') problem completely generally, predicting that typical observers find themselves at the onset of vacuum domination.  These claims may seem strong, so let us verify them by reviewing the basic argument leading to Eq.~(\ref{eq-ltobs}), and demonstrating that it is independent of the nature of observers.

In vacua with small cosmological constant ($\Lambda\ll t_{\rm obs} ^{-2}$), vacuum energy is dynamically irrelevant at and before the time $t_{\rm obs}$ and so, in particular, cannot affect the number of observers.  In this range of $\Lambda$, the probability distribution is governed by the prior landscape distribution, which grows towards larger values of $\Lambda$, like $\Lambda$.  In vacua with large cosmological constant ($\Lambda\gg t_{\rm obs} ^{-2}$), the number of observers in the causal diamond is {\em at least\/} exponentially suppressed due to the depletion of the total mass by the de~Sitter expansion, like $\exp(-\sqrt{3\Lambda}\, t_{\rm obs})$~\cite{BouHar07}.  In this range, the probability distribution favors smaller values of $\Lambda$, so the overall distribution peaks near $\Lambda\sim t_{\rm obs} ^{-2}$.  Now, it is possible that sufficiently large values of the cosmological constant disrupt the formation of observers altogether (e.g., for observers like us, by disrupting galaxy formation).  Whether this happens, and for what values of $\Lambda$, will indeed depend on the type of observers and the dynamical processes that lead to their formation.   But any such additional suppression can only occur if $\Lambda$ has some dynamical effect before $t_{\rm obs}$, i.e., for values of $\Lambda$ greater than $t_{\rm obs}^{-2}$.  Since the distribution is already suppressed in this regime, this will not significantly affect the prediction, Eq.~(\ref{eq-ltobs}).

Let us summarize the two reasons why Eq.~(\ref{eq-ltobs}) is a very general prediction: (1) both $\Lambda$ and $t_{\rm obs}$ are well-defined observable quantities in any semi-classical vacuum---unlike, for example, the number of galaxies or carbon atoms; and (2), no additional vacuum-specific assumptions went into deriving their correlation, Eq.~(\ref{eq-ltobs}); in particular, no assumption about the nature of the observers was made, and no necessary condition for their existence was posited.  Unfortunately, examination of the argument in the previous paragraph suggests that this fortuitous circumstance is quite peculiar to the pair of variables appearing in Eq.~(\ref{eq-ltobs}). Perhaps a few other examples can be found.  But short of demonstrating that a certain property cannot occur at all in the entire vacuum landscape~\cite{Vaf05}, it does not seem likely that many predictions can be made without assuming {\em anything\/} about observers.  

This does not mean that we must give up on exploring the whole landscape.  Rather, the challenge is to identify a universal necessary condition for observers: one that is plausible but not so specific as to restrict the class of observers, or the class of vacua, that can be considered.  Galaxies or carbon may be good proxies for observers in restricted classes of vacua, with their abundance tracking the true number of observers reasonably well.  But they will not do for this more ambitious purpose.  

\subsection{The entropic principle}

In Ref.~\cite{Bou06}, a novel proxy for observers was proposed: the production of matter entropy.  The formation of any complex structure (and thus, of observers in particular), is necessarily accompanied by an increase in the entropy of the environment.  Thus, entropy production is a necessary condition for the existence of observers.  In a region where no entropy is produced, no observers can form.  The ``entropic principle'' is the assumption that {\em the number of observers is proportional, on average, to the amount of entropy produced}:
\begin{equation}
n_{\rm obs}\propto \Delta S~.
\end{equation}

Entropy production is obviously not {\em sufficient\/} for the existence of observers, just as galaxies are not sufficient.  The entropic principle makes no claim of establishing an equivalence where previously only a necessary condition was known.  Rather, its potential advantage lies in the universality of the notion of entropy.  A galaxy, or an organic molecule, is an imprecise and highly derivative concept with no obvious analogue in vacua with different cosmology, particles or forces.  Matter entropy, by contrast, is defined in any vacuum that admits a semi-classical description.\footnote{In particular, matter entropy can be distinguished from the Bekenstein-Hawking entropy associated with black hole or cosmological horizons, in the semi-classical regime.  We do not include horizon entropy in $\Delta S$.  This is legitimate, since we are not claiming that entropy production is fundamentally equivalent to observation; we are merely interested in exploring matter entropy production as a proxy for observers that can be defined in all vacua.  However, it is worth noting that the distinction between matter and horizon entropy is particularly natural in the context of the causal patch and causal diamond measures.  The cosmological horizon and the horizons of all black holes are by definition on or just outside the boundary of the causal patch, and so are automatically excluded from consideration. \label{fn}} Therefore, its use as a proxy does not require us to restrict our attention to a small subset of the landscape.

Before applying the entropic principle on the whole landscape (the purpose of the present paper), let us check whether entropy production makes for a good proxy in vacua like ours, where we have alternative, more conventional methods of estimating the abundance and spacetime distribution of observers.  Beginning with our own vacuum, let us pretend, for a moment, that we do not already know our position in space and time, and use the entropic principle to ``predict'' our location.  Analysis of a variety of known astrophysical processes~\cite{BouHar07} shows that the entropy production in the causal diamond in our universe is dominated by the infrared photons emitted by dust heated by starlight.  According to the entropic principle, this implies that typical observers should find themselves near stars in galaxies (and not, for example, in a more generic location like intergalactic space).  The rate of entropy production peaks at a time of order $10^{10}$ years~\cite{BouHar07}, so the entropic principle predicts that we should find ourselves living at about that time.  Both of these predictions are correct, which suggests that the entropy production is indeed a useful proxy for observers.  

Next, consider an ensemble of vacua that includes our own, as well as (a) vacua like ours except without complex molecules, (b) vacua like ours but without stars, and (c) vacua like ours but without galaxies.  In our vacuum, dust scatters a fraction of order one of stellar photons, converting each optical photon into about 100 infrared photons.  Vacua of type (a) do not contain dust, which is composed of large organic molecules~\cite{andriesse,draine}.  Therefore, the entropy production in vacua without organic chemistry is about 100 times smaller than in our vacuum.  In vacua without stars, the entropy is even smaller (dominated, perhaps, by AGNs, if those still exist, or by galaxy cooling); and without galaxies, $\Delta S$ is smaller still.  This shows that $\Delta S$ is remarkably sensitive to some of the rather specific criteria that have been conjectured to be necessary for life of our type.  This further boosts our confidence that it will be a useful proxy in the larger landscape.

\subsection{Outline}

In this paper, we will study the implications of the entropic principle applied to the landscape as a whole.  The basic idea is simple: if entropy production traces observers, then most observers should find themselves in an environment in which a lot of entropy is produced.  The question is how to turn this insight it into quantitative predictions of observable quantities.  Our approach will be to define a set of observables $x_i$ that characterize processes that produce entropy, and to study how $\Delta S$ depends on the parameters $x_i$.  By the entropic principle, one expects that most observers will be witness to a process that comes close to maximizing the amount of entropy production.  Thus, we predict that the observed values of the parameters $x_i$ are not far from the values that maximize $\Delta S$.

In Sec.~\ref{sec-parameters}, we define six scanning parameters: the time of entropy production, $t_{\rm burn}$; the characteristic wavelength of the quanta carrying the entropy, $\lambda$; the volume fraction occupied by the quanta, $f_V$, which is small if quanta are dilute; the characteristic occupation number, $\eta$, which quantifies the typical overlap between entropic quanta; the spatial curvature, parametrized by the time of curvature domination, $t_c$, and the rest mass of entropic quanta, $m$.  In order to be able to judge the sharpness of the predictions we will later derive from the causal entropic principle, we discuss the considerable range of values that can be taken in principle by each parameter.  

In Sec.~\ref{sec-main}, we consider a (nearly) arbitrary entropy production process: free energy is converted into (nearly) arbitrary quanta, which we refer to as ``entropic quanta''.  We consider a vacuum with cosmological constant $\Lambda$, which we treat as an input parameter much like $t_{\rm obs}$ in the example of Eq.~(\ref{eq-ltobs}).  By maximizing the entropy produced, $\Delta S$, over our scanning parameters, we obtain predictions for all six parameters.

In Sec.~\ref{sec-examples}, we consider concrete examples processes that produce entropy.   This demonstrates explicitly that entropy production is smaller than the maximal value in universes that burn up too little of their rest mass, or produce only classical radiation, or produce thermal radiation at a temperature that is higher or lower than optimal.

\begin{figure}[t]
\begin{center}
\includegraphics[height=0.49\textwidth]{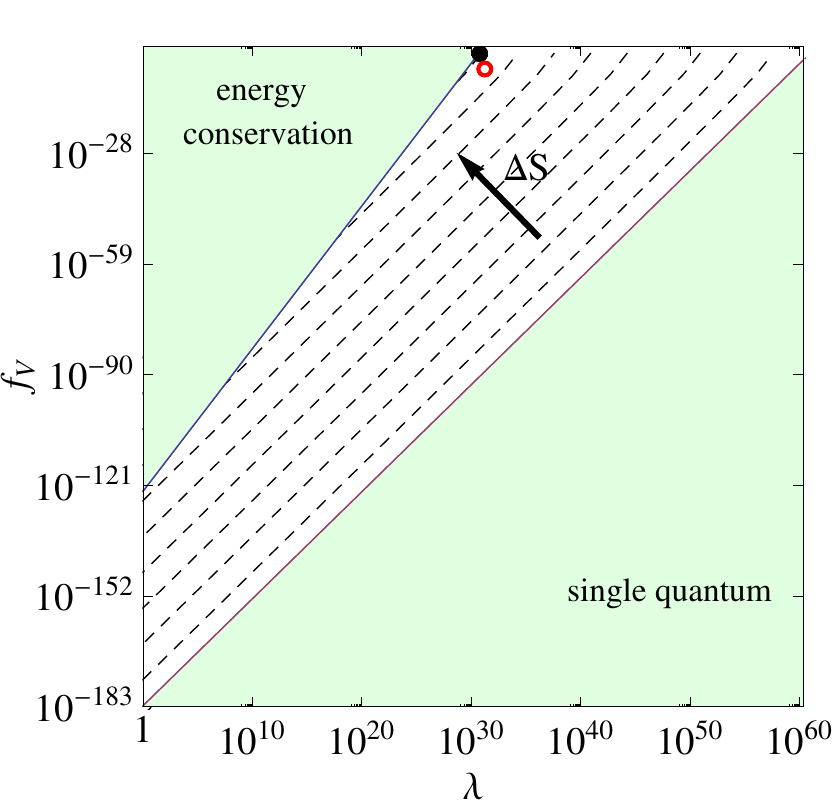}
\includegraphics[height=0.49\textwidth]{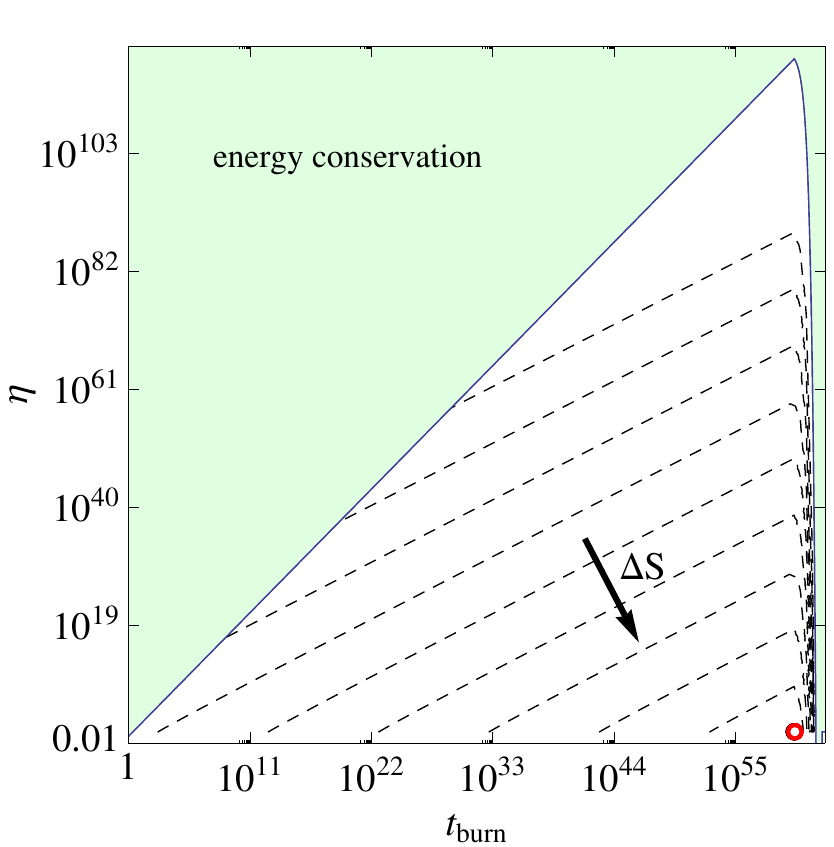}
\caption{
Two slices of our parameter space: the ($\lambda$, $f_V$) plane on the left,
and the ($t_{\rm burn}$, $\eta$) plane on the right. In each plot, the remaining parameters are held fixed at their optimal values. The axes show the possible values these parameters may take (in Planck units), in vacua with our value of the cosmological constant. The shaded regions are unphysical, either because the energy of the entropic radiation exceeds the total mass within the causal diamond, or because there is less than one quantum of radiation in the diamond.  The dashed contours are surfaces of constant entropy (ten orders of magnitude apart in entropy) and the black arrow represents the direction in which entropy production increases.  The optimal values, for which entropy production is maximized, are shown as a black dot.  The observed values in our Universe are shown as a red open circle; as predicted by the entropic principle, they are very close to the optimal values.  (The black dot cannot be seen in the figure on the right because it is covered by the red circle.)
}
\label{predictionplot}
\end{center}
\end{figure}

In Sec.~\ref{sec-compare}, we turn to our own universe and compare our predictions to the values that we observe for the scanning parameters $t_{\rm burn}, \lambda, f_V, \eta, t_c, m$.
(We briefly summarize our predictions, and how they compare to observation, in Sec.~\ref{sec-summary} immediately below.)

We finish in Sec.~\ref{sec-discussion} by discussing our results and considering future extensions.  We sketch a statistical explanation of the near-coincidence between the temperature of the cosmic microwave background and the cosmic infrared background, for which no natural explanation is known.  We also consider what happens if $\Lambda$ is allowed to scan: we argue that the observed value of the cosmological constant may be directly related to the number of vacua in the landscape, providing a fundamental hierarchy to which other known hierarchies are correlated.

The appendix contains a brief review of the definition of the causal diamond measure.

\subsection{Summary of results}
\label{sec-summary}

Let us briefly summarize our predictions and how they compare to observation (see Fig.~\ref{predictionplot}): 
\begin{enumerate}

\item {\em Prediction:} $t_{\rm burn} \sim t_\Lambda\equiv |3/\Lambda|^{1/2}$:  Most of the entropy in the causal diamond is produced around the onset of vacuum domination at the time $t_\Lambda$.   By the entropic principle, this is equivalent to the prediction that observers should find themselves at the onset of vacuum domination, and thus, to Eq.~(\ref{eq-ltobs}).\\[1ex] {\em Observation:} The dominant entropy production processes that we observe or infer from observation take place around $3.5\, \mbox{Gyr} \sim 10^{60.3}$ after the big bang~\cite{BouHar07}.  (Of these, the most important is the emission of infrared radiation by galactic dust.)  In our universe, $t_\Lambda \approx 10^{60.3}$, so the prediction is accurate.

\item  {\em Prediction:} $\lambda\sim |\Lambda|^{-1/4}$: The characteristic energy of the quanta of matter carrying this entropy is of order the energy scale of the vacuum. \\[1ex] {\em Observation:} The energy of quanta emitted by galactic dust is observed to be about  300$\mu$m$\sim 10^{31.3}$, which is indeed of order $\Lambda^{-1/4}\sim 10^{30.8}$.

\item  {\em Prediction:} $\eta \sim 1$: The quanta produced by this process are dilute, in the sense that occupation numbers are of order unity.  The quanta are not in classical radiation, which would have $\eta\gg 1$. \\[1ex] {\em Observation:} Experimentally, we know that dust will not give off classical radiation when scattering optical photons, so $\eta\sim 1$.

\item  {\em Prediction:} $f_V\sim 1$: Entropic quanta are nearly dense in space.  Therefore, they will be observed as a steady flux of radiation, rather than as bursts, and they will appear to come from a large fraction of the sky. 
\\[1ex] {\em Observation:}  
We observe the glow of galactic dust from the Universe as a roughly isotropic cosmic infrared background (CIB) which is steady to within the time resolution of current detectors. Given the flux and wavelength of the CIB we can however deduce that it is somewhat dilute, with an $f_V\sim 0.5\times 10^{-4}$. Given that $f_V$ may a priori span over 180 orders of magnitude, this value is remarkably close to 1.

\item  {\em Prediction:} The universe is not curvature dominated, i.e., the radius of spatial curvature is larger than the Hubble radius. \\[1ex] {\em Observation:} Curvature does not dominate.  The curvature radius is at least 10 times larger than the Hubble radius~\cite{WMAP5}.

\item  {\em Prediction:} The quanta produced by this process are relativistic. \\[1ex] {\em Observation:} The cosmic infrared background consists of photons, i.e., of massless particles, which are obviously relativistic.

\end{enumerate} 

Let us give a rough explanation of how the entropic principle, combined with the causal diamond cut-off, leads to the above predictions.  The causal diamond contains the most mass (which is an upper bound on free energy) at the time $t_\Lambda$, so this is the best era in which to produce entropy by burning up free energy.  For a dilute gas of quanta, the entropy is proportional to the number of quanta produced, which is given by the free energy divided by the energy per quantum: $S\sim N\sim M\lambda$.  To maximize the entropy, therefore, one would like to make the wavelength $\lambda$ large.  But this also increases the volume occupied by each quantum.  The total volume the quanta can occupy is limited by causality, so for $\lambda$ sufficiently large the quanta can no longer be dilute, but would have to overlap ($\eta\gg 1$).   This happens when $\lambda\sim \Lambda^{-1/4}$.  Increasing the number of quanta beyond this point is disfavorable to entropy production.  (This is most obvious in the extreme limit where $\eta\sim N$ and $S=\log N$.)   Thus, the entropy is maximized when the photons make use of all available volume, $f_V\sim 1$ while remaining marginally dilute, $\eta\sim 1$.  Curvature is disfavored because it decreases the maximum volume available to the entropic radiation.  Non-relativistic quanta are disfavored because some of the mass inside the causal diamond goes into the rest mass of particles, decreasing the number of quanta that can be produced.

\section{Parameters and their ranges}
\label{sec-parameters}

Consider a Friedmann-Robertson-Walker (FRW) universe with cosmological constant $\Lambda$, which is taken to be a fixed input parameter.  We will now define six scanning parameters $(t_{\rm burn}, \lambda, f_V, \eta, t_c, m)$, which can be used to characterize entropy production in arbitrary vacua, and whose values we will later predict in terms of $\Lambda$.

Entropy is produced by an arbitrary process, during an era $t_{\rm burn}$, in the form of quanta of rest mass $m$, with characteristic wavelength $\lambda$ per quantum.  We will refer to the carriers of this entropy as {\em entropic quanta\/} or collectively as {\em entropic radiation}.   In the following discussion, we will neglect factors of order unity.   We define $f_V$ to be the fraction of volume occupied by entropic radiation.  We define $\eta$ as the characteristic occupation number of entropic quanta, {\em in the regions occupied by at least one such quantum}. (One could also consider an occupation number averaged over all space, which would be $f_V\eta$ in our notation.  However, it will be more convenient to work with the above definition.) Finally, we define $t_c$ as the time when (open or closed) spatial curvature would begin to dominate the evolution of the universe if the cosmological constant were zero.  That is $t_c$ is the actual time of curvature domination if $t_c<t_\Lambda\sim \Lambda^{-1/2}$; for $t_{\rm c}>t_\Lambda$, curvature never dominates after reheating, and the universe can be considered spatially flat on the scale of the causal diamond.

Let us examine the range of values each scanning parameter can take.   The time $t_{\rm burn}$ cannot be smaller than the Planck time. One might think that it can take arbitrarily large values.  However, after a time $t_\Lambda\equiv (\Lambda/3)^{-1/2}$, the universe begins to accelerate its expansion.  As matter exits from the de~Sitter horizon, the total mass of matter in the causal diamond drops like $t_\Lambda \exp(-3t/t_\Lambda)$.  This becomes less than the smallest reasonable quantity of mass that can be defined, a quantum of wavelength $t_\Lambda$, when $\exp(3t/t_\Lambda)\sim t_\Lambda^2$.  Thus, we find that
\begin{equation}
1 \leq  t_{\rm burn} \leq  \frac{2}{3} t_\Lambda \log t_\Lambda ~.
\end{equation}
For example, in vacua with our value of the cosmological constant, $\Lambda=\Lambda_o=3.72\times 10^{-122}$, this implies that $t_{\rm burn}$ can range over 63 orders of magnitude.  (In section~\ref{sec-nonhomogeneity} we will address nonhomogeneities that can in principle make $t_\mathrm{burn}$ even larger.)

The wavelength of entropic quanta cannot be smaller than the Planck length or larger than the de~Sitter horizon:
\begin{equation}
1 \leq  \lambda \leq  t_\Lambda~;
\end{equation}
with $\Lambda =\Lambda_o$, this implies that $\lambda$ can range over 61 orders of magnitude.  

The volume fraction occupied by radiation, $f_V$, cannot exceed 1 (by definition).  The smallest it can be is $1/V$, where $V$ is the total volume available; this corresponds to a single quantum of Planck size.  The maximum volume of the causal diamond in a universe with cosmological constant $\Lambda$ is of order $t_\Lambda ^3$, so
\begin{equation}
t_\Lambda^{-3}\leq f_V\leq 1~.
\label{eq-fvrange}
\end{equation}
This corresponds to a range of 183 orders of magnitude in vacua with $\Lambda=\Lambda_o$.

The typical occupation number in regions containing entropic radiation, $\eta$, cannot be less than 1 by definition; and it cannot be larger than the total number of quanta, $N$.   As we will discuss in more detail in Sec.~\ref{sec-main}, the total mass inside the causal diamond is maximal around the time $t_\Lambda$, when it is given by $t_\Lambda$.  The smallest possible mass of a single quantum is $t_\Lambda^{-1}$ (which is attained when its wavelength is as large as the universe).  The largest possible $N$ corresponds to dividing the maximum total mass in the universe by the energy of the smallest possible quanta.   This yields $N\leq t_\Lambda^2$ and corresponds to a condensate of $\eta=N$ quanta, each of horizon size, sitting on top of one another.  Thus,
\begin{equation}
1\leq \eta\leq t_\Lambda^2~,
\end{equation}
corresponding to a range of 122 orders of magnitude in vacua with $\Lambda=\Lambda_o$.

The rest mass of the species carrying the entropic radiation can range from the Planck mass down to zero mass, and so has infinite range in principle.  In practice, it is unclear how one would measure the mass of a particle whose Compton wavelength exceeds the largest causally accessible scale, $t_\Lambda$, so we take the mass range to be
\begin{equation}
m=0 ~~\mbox{or} ~~t_\Lambda^{-1}\leq m\leq 1~.
\end{equation}
This corresponds to 61 orders of magnitude in vacua with $\Lambda=\Lambda_o$ (plus the special case of massless particles, which we shall see is actually the main carrier of entropic radiation in our universe).

The curvature time scale, $t_c$, ranges in practice from the time of reheating (which could be as early as the Planck time) to the time of vacuum domination, $t_\Lambda$; larger values of $t_c$ correspond to a universe which always appears flat on the horizon scale.  Thus,
\begin{equation}
1\leq t_{\rm c} \leq t_\Lambda ~,
\end{equation}
corresponding to a range of 61 orders of magnitude in vacua with $\Lambda=\Lambda_o$.

We summarize our parameters and their range in the following table~\ref{tab-parameters}:
\begin{table}[h]
\begin{center}
\begin{tabular}{|c|c|c|}
\hline
parameter & description & $\log_{10}$(allowed range) \\
\hline
$t_\mathrm{burn}$ & time of entropy production &  63      \\
$\lambda$ & characteristic wavelength of radiation &  61      \\
$f_V$ & volume fraction take up by radiation ($\leq 1$) &  183    \\
$\eta$ & typical \# of quanta that overlap each other ($\ge1$) &  122      \\
$m$ & rest mass of the radiation species &  61      \\
$t_c$ & time of curvature domination &  61      \\
\hline
\end{tabular}
\caption{A summary of our parameters.  The last column shows the number of orders of magnitude each parameter can span in vacua with our value of the cosmological constant.}\label{tab-parameters}
\end{center}
\end{table}
Thus we are considering a six-dimensional parameter space, with each parameter allowed an enormous range {\em a priori}.  This should be kept in mind in order to appreciate the accuracy of the predictions of the entropic principle, which we derive next.  Note that each point in the parameter space can correspond to a huge variety of different entropy production processes.   We have not specified how the entropic radiation is actually generated, so that our analysis remains sufficiently general to apply to all semi-classical landscape vacua. 

\section{Entropy and its maximization}
\label{sec-main}

\subsection{The entropy of quanta in a grid}
\label{sec-grid}

The standard method of doing statistical mechanics of a noninteracting gas of particles is to compute the expected occupation numbers of the one-particle energy eigenstates, or ``orbitals'', in a box.  These numbers are interpreted as the number of photons (or quanta of some other field) of the corresponding energy.  These states, however, have support over the entire volume of the box and are (by definition) exactly time-independent.  One of the key observables studied in this paper is the volume fraction occupied by the radiation, so this basis is not well-suited to our purposes.  Therefore, we will use a basis of one-particle states which are not energy eigenstates but are localized in space.  At a given characteristic wavelength (whose variation we will consider later) we may consider a grid consisting of $c$ cells of size $\lambda^3$.  The volume fraction defined in the previous section can be represented as 
\begin{equation}
f_V\equiv \frac{d}{c}
\label{eq-fdc}
\end{equation}
where $d$ is the number of cells containing at least one quantum (``occupied'' cells).  The characteristic occupation number is
\begin{equation}
\eta\equiv \frac{N}{d}~.
\label{eq-end}
\end{equation}

Given $N$, $c$, and $d$ (or equivalently, $N$, $f_V$ and $\eta$), the entropy is 
\begin{equation}
S=\log  ({\cal N}_f \, {\cal N}_\eta)~,
\end{equation}
where
\begin{equation}
{\cal N}_f=\left(\begin{array}{c} c\\ d\end{array}\right)
\label{eq-nf}
\end{equation} 
is the number of different ways of choosing the $d$ occupied cells from among the $c$ cells, and
\begin{equation}
{\cal N}_\eta=\left(\begin{array}{c} N-1\\ d-1\end{array}\right)
\label{eq-neta}
\end{equation}
is the number of different ways of distributing $N-d$ quanta among the $d$ occupied cells. (By definition, each occupied cell contains at least one quantum, so only the remaining $N-d$ quanta are left to be distributed freely.)  We will be assuming that $N$, $c$, and $d$ are all much greater than one, so that the $-1$ terms can be neglected in Eq.~(\ref{eq-neta}).

\subsection{Optimizing the distribution of quanta}
\label{sec-dist}

While holding fixed the number of quanta, $N$, and the number of cells, $c$, let us ask for what value of $d$ the entropy is maximized.  From Eqs.~(\ref{eq-nf}) and (\ref{eq-neta}), this occurs when ${\cal N}(d+1)={\cal N}(d)$, or 
\begin{equation}
d=\frac{Nc}{N+c}~,
\label{eq-dopt}
\end{equation}
where we have neglected terms of order unity both in the numerator and denominator. As one would intuitively guess, entropy is maximized when the number of occupied cells $d$ is as large as possible, set either by the number of cells $c$ or by the number of quanta $N$, whichever is smaller. 
After picking the optimal value for $d$, the entropy is
\begin{equation}
S= c \log\left( \frac{N+c}{N}\right) + N \log\left( \frac{N+c}{c}\right)\,.
\label{eq-scn}
\end{equation}
where we have employed the Sterling formula on large factorials. 

So far, we were able to analyze the entropy as a purely combinatorial problem with no scales involved. At this level, increasing the number of quanta and/or the number of cells is beneficial for increasing the entropy. However, in the next step we will account for the limited resources, energy and volume, available in a physical situation. The dimensionless parameters $N$ and $c$ may be traded for ratios of dimensionful physical parameters (two ratios will require three scales). 

\subsection{Optimizing the wavelength in a physical system}
\label{sec-wavelength}

Consider a physical system of volume $V$ and a certain amount of energy $M$ that is in the form of radiation. Let use consider only quanta of a certain wave length $\lambda$. 
The number of quanta that may be produced is the total mass, divided by $\lambda^{-1}$, the energy of a single quantum (here and below we drop factors of order unity):
\begin{equation}
%N=\frac{M\lambda}{2\pi}\,.
N= M\lambda\,.
\label{eq-nml}
\end{equation}
The number of cells of volume $\lambda^3$ that fit within a certain volume $V$ is
\begin{equation}
%c=\frac{4\pi V}{\lambda^3}\,.
c= \frac{V}{\lambda^3}\,.
\label{eq-cvl}
\end{equation}
In this subsection we will hold $M$ and $V$ fixed and vary only the wavelength $\lambda$. In terms of the parameters of the previous subsection, we are varying $N/c$ while keeping $N^3c$ fixed to the value $M^3V$.
Now, for a fixed mass and volume, increasing the wavelength will increase the number of quanta but decrease the number of cells. Increasing both in order to get an arbitrarily large entropy is no longer possible.   
By Eq.~(\ref{eq-scn}), the entropy is
\begin{equation}
%S= \frac{4\pi V}{\lambda^3} 
%\log\left(1+ \frac{ 16\pi^2 V}{M\lambda^4}\right) 
%+ \frac{M\lambda}{2\pi} \log\left(1+  \frac{M\lambda^4}{16\pi^2 V}\right)\,.
S= \frac{V}{\lambda^3} 
\log\left(1+ \frac{V}{M\lambda^4}\right) 
+ M\lambda\log\left(1+  \frac{M\lambda^4}{V}\right)\,.
\label{eq-svlm}
\end{equation}
the entropy is maximized when the two terms contribute equally, so the optimal wavelength is
\begin{equation}
\lambda \to \left( \frac{M}{V} \right)^{-1/4}~.
\label{eq-optwavelength}
\end{equation}
In terms of the parameters of the previous subsection, Eq.~(\ref{eq-optwavelength} is equivalent to 
\begin{equation}
\frac{N}{c}\sim 1 \qquad \mbox{or} \qquad N\sim c
\end{equation}
Combining this result with Eq.~(\ref{eq-dopt}), we find 
\begin{equation}
d\sim c \sim N
\end{equation}
for maximum entropy.  
The maximal entropy that may be produced is
\begin{equation}
\label{eq-maxent}
S\sim N \sim M^{3/4} V^{1/4}~,
\end{equation}

Before continuing to determine the remaining two parameters, we will pause to express the predictions we have made thus far in terms of two of the scanning parameters introduced in section~\ref{sec-parameters},
\begin{equation}
\eta=\frac{N}{d} \to 1 \qquad \mbox{and} \qquad f_V=\frac{d}{c} \to 1~.
\end{equation}
Thus we have shown  that an optimal entropy producing system should be nearly filled with non-overlapping quanta.
We have also shown that in order to maximize the entropy in a system of volume $V$ and total energy $M$, the wavelength of the radiation should obey $\lambda^4\sim V/M$. We will now embed this system into a causal diamond set by a cosmological constant $\Lambda$ at a particular time $t_\mathrm{burn}$. Optimizing over $M$ and $V$ will allow us to make predictions for $\lambda$ and $t_\mathrm{burn}$ in terms of $\Lambda$ alone.

\subsection{Optimizing the energy and volume in the causal diamond}
\label{sec-envol}

In a flat FRW universe with positive cosmological constant $\Lambda$,\footnote{Generalizations are considered in Sec.~\ref{sec-generalizations}.} the mass inside the causal diamond is given by
\begin{equation}
\label{eq-diamondmass}
M_{\rm CD}(t)\sim \left\{
\begin{array}{lll}
t & \mbox{for} & t<t_\Lambda \\
t_{\Lambda} e^{-3(t/t_\Lambda-1)} & \mbox{for} & t>t_\Lambda
\end{array} \right.
\end{equation}
A process taking place at the time $t_{\rm burn}$ (which we still hold fixed for now) will maximize entropy production by converting all of this mass into radiation.  Thus, the entropic principle predicts:
\begin{equation}
M\to M_{\rm CD}(t_{\rm burn})~.
\end{equation}
Our final step is to maximize the resulting entropy
\begin{equation}
\Delta S\sim M_{\rm CD}(t_{\rm burn})^{3/4} V(t_{\rm burn})^{1/4}
\label{eq-smv}
\end{equation}
over $t_{\rm burn}$.  

The physical volume of the causal diamond is given by
\begin{equation}
V_{\rm CD}(t)\sim  \left\{
\begin{array}{ll}
t^3 &  \mbox{for }t<t_\Lambda \\
t_{\Lambda}^3 & \mbox{for }t>t_\Lambda~
\end{array} \right.~.
\label{eq-diamondvolume}
\end{equation}
The actual volume available for entropic radiation produced in the ``bottom cone'' of the causal diamond is precisely equal to the volume of the causal diamond itself
\begin{equation}
V=V_{\rm CD} ~~\mbox{for}~~t_{\rm burn} \lesssim t_\Lambda ~.    
\end{equation}
The volume available to quanta produced at late times,
\begin{equation}
V\sim t_\Lambda^3 \exp\left(\frac{3 t_{\rm burn}}{t_\Lambda}\right) ~~\mbox{for}~~t_{\rm burn} \gg t_\Lambda ~.    
\end{equation}
is much larger because quanta are stretched outside the horizon by the de~Sitter expansion. Substituting into Eq.~(\ref{eq-smv}) one finds
\begin{equation}
\Delta S\sim \left\{
\begin{array}{lll}
t_{\rm burn}^{3/2} &  \mbox{for} & t_{\rm burn} \lesssim t_\Lambda \\
t_{\Lambda}^{3/2}\exp\left(-\frac{3}{2}\frac{t_{\rm burn}}{t_\Lambda}\right) & \mbox{for} & t_{\rm burn} \gg t_\Lambda~
\end{array} \right. ~.
\end{equation}
This is maximal for
\begin{equation}
t_{\rm burn} \to t_\Lambda~.
\end{equation}
Picking $t_{\rm burn}$ sets the total energy and volume occupied by radiation to
\begin{equation}
M \to \Lambda^{-1/2} \qquad\mbox{and}\qquad V\to \Lambda^{-3/2}
\end{equation}
Using equation~\ref{eq-optwavelength} for the optimal wavelength we get 
\begin{equation}
\lambda\to \Lambda^{-1/4}~.
\end{equation}

Collecting our results, we find that entropy production is maximized for
\begin{eqnarray}
t_{\rm burn} & \to & \Lambda^{-1/2}~;
\label{eq-tb}\\
\eta & \to & 1~;
\label{eq-eta}\\
f_V & \to  & 1~;
\label{eq-v}\\
\lambda & \to  & \Lambda^{-1/4}~,
\label{eq-temp}
\end{eqnarray} 
We predict that the observed values of these four parameters should not be many orders of magnitude from these optimal values.

At the optimal values, the entropy produced is
\begin{equation}
\Delta S_{\rm max} \sim \Lambda^{-3/4}~,
\label{eq-smax}
\end{equation}
We should make it clear that this result for $\Delta S_{\rm max}$ is only a corollary to our main result, Eqs.~(\ref{eq-tb})-(\ref{eq-temp}).  Our goal was to predict several directly observable parameters by expressing $\Delta S$ as a function of these parameters and maximizing.\footnote{Indeed, Eq.~(\ref{eq-smax}) should come as no surprise.  In matter- or radiation-dominated cosmologies, the causal diamond coincides (up to factors of order unity) with the interior of the apparent horizon---roughly, the radius out to which matter inside the universe can be confined by some force and rendered stationary.  Stationary matter configurations confined to a spherical region of area $A$ satisfy $S\leq A^{3/4}$~\cite{Tho93}.  With $A\sim\Lambda^{-1}$ for the maximum area of the apparent horizon in de~Sitter space, Eq.~(\ref{eq-smax}) follows.}

\subsection{Generalizations}
\label{sec-generalizations}

We have considered a very general entropy production process that might take place in a vacuum completely unlike our own.  Nevertheless, we did make a few assumptions in order to keep the analysis simple.  We will now eliminate some of these assumptions, and we will argue that our results are unchanged. 

In two cases, we will go further and show that the assumptions themselves can be obtained as predictions from the causal entropic principle, because they turn out to be necessary for maximizing entropy production.  That is, we shall find that the production of nonrelativistic quanta of entropy, or entropy production in spatially curved universes, is necessarily suboptimal.  This not only eliminates the assumptions of relativistic quanta and spatial flatness, but recovers them as predictions of the causal entropic principle.

\subsubsection{Entropy in nonrelativistic species}
\label{sec-nonrel}

In Sec.~\ref{sec-wavelength}, we considered the production of entropy in {\em relativistic\/} quanta.  We will now lift this restriction and show that it is recovered as a prediction.  This is because in the nonrelativistic case, most of the available energy goes into the rest mass of the entropic particles and is unavailable for increasing the number of particles.

Consider a process by which $N$ quanta of an arbitrary, relativistic or nonrelativistic, species are produced.  Then the number of quanta is given by
\begin{equation}
N=\frac{M}{(\lambda^{-2}+m^2)^{1/2}}~,
\end{equation}
where $m$ is the mass of the entropic species.  This reduces to Eq.~(\ref{eq-nml}) in the relativistic case, $m\ll \lambda^{-1}$.

In the nonrelativistic case, $m\gg \lambda^{-1}$, an analysis analogous to that of Secs.~\ref{sec-wavelength} and~\ref{sec-envol} shows that entropy production is maximized for
\begin{eqnarray}
t_{\rm burn} & \to & \Lambda^{-1/2}~;
\label{eq-tbnr}\\
\eta & \to & 1~;
\label{eq-etanr}\\
f_V & \to  & 1~;
\label{eq-vnr}\\
\lambda & \to  & \frac{m^{1/3}}{\Lambda^{1/3}}~.
\label{eq-tempnr}
\end{eqnarray} 
The entropy produced is 
\begin{equation}
\Delta S_{\rm NR}\sim\frac{t_\Lambda}{m}=\Lambda^{-1/2} m^{-1}~.
\end{equation}
The nonrelativistic assumption implies $m\gg \Lambda^{1/4}$, which in turn implies that
\begin{equation}
\Delta S_{\rm NR}\ll \Lambda^{-3/4}~.
\end{equation}
Thus, the maximum entropy that can be produced in nonrelativistic species is necessarily smaller than the maximum entropy that can be produced in relativistic species, Eq.~(\ref{eq-smax}).  The entropic principle {\em predicts\/} that entropy will be dominantly produced as relativistic particles.

\subsubsection{Spatial curvature}

In Sec.~\ref{sec-envol}, we computed the maximum amount of entropy that can be produced in a spatially flat Friedmann-Robertson-Walker universe.  We will now show that this amount, $\Delta S_{\rm max}$, cannot be further increased by allowing for positive or negative spatial curvature.

If curvature fails to dominate before $t_\Lambda $, then it will never play a dynamical role, and the analysis of the previous subsections remains unchanged.  If it begins to dominate around the time $t_\Lambda $, then $\Delta S_{\rm max}$ is changed only by a factor of order unity.  Thus, we need only consider the case $t_{\rm c}\ll t_\Lambda $, where $t_{\rm c}$ is the time of the onset of curvature domination over matter.

For positive curvature, the universe recollapses on a timescale of order $t_{\rm c}$.  The tip of the causal diamond will be at the big crunch, and the edge will be at the time of maximum expansion.  The maximum entropy production can be estimated, up to order one factors, by imposing the additional constraint $t\lesssim t_{\rm c}$ in the analysis of Sec.~\ref{sec-envol}.  (In particular, the case $t>t_\Lambda$ does not arise then.)  With this restriction, entropy production is maximized for $f_V\to 1$, $\eta\to 1$, $t_{\rm burn} \to t_{\rm c}$ and $\lambda\to t_{\rm c} ^{1/2}$, yielding $\Delta S_{\rm max}\sim t_{\rm c}^{3/2}$, which is much less than $\Lambda^{-3/4}\sim t_\Lambda^{3/2}$.  

For negative curvature, too, we need only reconsider the last subsection, where we maximize the entropy over the energy and volume available in the causal diamond.  Using the results of Ref.~\cite{BouLei09} (Sec.~5.2) for the geometry of the causal diamond, we find again that the optimal entropy production time, $t_{\rm burn}$, is at the ``edge'' time, when the upper and lower light-cone bounding the causal diamond intersect.  However, in the presence of negative curvature this time is given by
\begin{equation}
t_{\rm edge} \sim (t_\Lambda t_{\rm c} )^{1/2}~,
\end{equation}
so we predict
\begin{equation}
t_{\rm burn} \to \Lambda^{-1/4} t_{\rm c}^{1/2} ~.
\label{eq-tburncurv}
\end{equation}
At this time, the upper bound on the energy of the radiation is still
\begin{equation}
M_{\rm CD} (t_{\rm burn})\sim t_\Lambda ~.
\end{equation}
but the available volume is smaller than in the flat case:
\begin{equation}
V_{\rm CD}(t_{\rm burn})\sim t_\Lambda^{1/2} t_{\rm c}^{5/2}~.
\end{equation}
Using Eq.~(\ref{eq-optwavelength}) we find that the entropy is maximized for
\begin{equation} 
\lambda \to t_{\rm c} ^{1/2}\left(\frac{t_c}{t_\Lambda}\right)^{1/8}~.
\label{eq-tempc}
\end{equation}
The maximum entropy produced is
\begin{equation}
\Delta S_{\rm max} \sim \Lambda^{-3/4} \left(\frac{t_{\rm c}}{t_\Lambda }\right)^{5/8}~,
\label{eq-smaxc}
\end{equation}
which for $t_{\rm c}\ll t_\Lambda $ is much less than the maximum entropy that can be produced in a flat universe, Eq.~(\ref{eq-smax}).

We conclude that both positive and negative spatial curvature decrease the maximum entropy that can be produced.  The entropic principle prefers a spatially flat universe.

\subsubsection{Nonhomogeneity}
\label{sec-nonhomogeneity}

An assumption of homogeneity appears to underlie Eq.~(\ref{eq-diamondmass}).  In a universe that is not perfectly, but only approximately homogeneous, like ours, this equation would appear to break down at sufficiently late times.  As all mass is expelled by the exponential expansion, the discreteness of the mass distribution implies that either one or zero bound structures remain inside the de~Sitter horizon.   

However, the causal diamond cutoff defines relative probabilities in terms of ratios of expectation values, which are computed from an ensemble of causal diamond reflecting the probabilities for different semiclassical histories.  At early times within a given vacuum bubble, this average is often dominated by a single history but when inhomogeneities are taken into account, then there are different possible histories at late times.  Some diamonds will be entirely empty at late times, while others contain a single gravitationally bound object.   However,  Eq.~(\ref{eq-diamondmass}) correctly describes the expectation value of $M(t)$ after averaging over all (empty and nonempty) de~Sitter horizon volumes at late times.  

Thus, a homogeneous mass distribution is not necessary for Eq.~(\ref{eq-diamondmass}). It suffices that the ensemble average reflects the distribution of empty and nonempty de~Sitter horizon volumes at late times.  This is what we assume here; whether it is actually the case depends on the precise definition of the ensemble. In the prescription given in Ref.~\cite{Bou06}, an ensemble of causal diamonds is generated by following timelike geodesics forward in time from fixed initial conditions.  It is unclear whether this leads to suitable averaging at late times~\cite{PhiAlb09}. Ultimately, it would be preferable to average over the future endpoints of geodesics (``terminal indecomposable past sets'' or tips), which are equivalent to causal patches~\cite{GarVil08,Bou09,BouYan09}.  One would expect this type of averaging to reflect the distribution of empty and nonempty Hubble volumes, but further study of the structure of future infinity of the multiverse will be needed to define and demonstrate this rigorously.

If Eq.~(\ref{eq-diamondmass}) breaks down, our conclusion may still be valid.  The entropy produced from a bound structure at late times can, in principle, be as large as $\Lambda^{-1}$, i.e., comparable to the horizon entropy of de~Sitter space~\cite{KraSta04}.  This is larger than the value we found for pre-vacuum-domination processes, $\Delta S_{\rm max}=\Lambda^{-3/4}$ in Eq.~(\ref{eq-smax}), by a factor $\Lambda^{-1/4}$. However, this process requires producing entropy at a temperature of order the Hawking temperature of the de~Sitter horizon.  It assumes that the matter structure persists for a time of order $t_\Lambda^2$ without disintegrating or forming a black hole.  To avoid $\eta\gg 1$, the process must produce no more than a single quantum every Hubble time, $t_\Lambda$.  This type of process is not continuously related to the kind of entropy production we see in our universe, and the dynamical probability for it to arise in a given vacuum may well be suppressed by more than $\Lambda ^{-1/4}$ relative to the processes that attain $\Delta S_{\rm max}=\Lambda^{-3/4}$ .

\subsubsection{Negative cosmological constant}
\label{sec-negative}

In Sec.~\ref{sec-main}, we used the cosmological constant, $\Lambda$, as a fixed input parameter.  We derived, as functions of this input parameter, the values of $(t_{\rm burn}, \eta, f_V, \lambda)$ that lead to optimal entropy production.  Note that we did not actually specify $\Lambda$, but we did assume that it was positive. In fact, however, the optimal  values of $(t_{\rm burn}, \eta, f_V, \lambda)$, and the associated maximum entropy production, $\Delta S_{\rm max}$, are independent of the sign of the cosmological constant.  That is, our predictions, Eqs.~(\ref{eq-tb}) -- (\ref{eq-temp}), hold more generally, with $\Lambda$ replaced by $|\Lambda|$.  This can be seen as follows.

For $\Lambda<0$, the analysis of Sec.~\ref{sec-grid} -- \ref{sec-wavelength} is completely unchanged, so we need only revisit the analysis of Sec.~\ref{sec-envol}.  The maximum mass inside the diamond is still attained at the time $t_{\rm edge}$, and we predict $t_{\rm burn} \sim t_{\rm edge}$.  With negative cosmological constant, the universe necessarily recollapses on a timescale of order $t_\Lambda\sim |\Lambda|^{-1/2}$.  (It may collapse sooner, if spatial curvature is large and positive, but we have already shown that this case is disfavored by the entropic principle.)  Thus, $t_{\rm edge}$ is at most of order $t_\Lambda$.  At early times, $t\ll t_\Lambda$, the cosmological constant is dynamically irrelevant, so the evolution is independent of the sign of $\Lambda$, and at the time $t_{\rm edge}\sim t_\Lambda$, the cosmological constant can affect the mass and volume of the causal diamond at most by a factor of order unity.  Thus, entropy production will be at best comparable to the positive $\Lambda$ case, but it cannot be made parametrically larger.

\section{Examples}
\label{sec-examples}

In the Sec.~\ref{sec-compare}, we will show that the predictions of the previous section agree well with the observed values in our universe.  The reader may object that this is trivial: all we have done is maximized entropy, which is what any system left alone long enough well tend to do.  In fact, it is not hard to imagine universes in which the matter entropy is very small compared to $\Lambda^{-3/4}$.   Moreover, we are interested in the entropy {\em produced}, not just the entropy present, within the causal diamond.  To illustrate these points, we will  consider some examples of entropy producing processes in this section.  This will also serve to illustrate the physical meaning of the parameters we have defined, and to give us a sense of which types of processes are effective generators of entropy.  

For definiteness, we will assume that the cosmological constant agrees with the value observed in our universe, $\Lambda=\Lambda_o\approx 3 \times 10^{-122}$.  In Secs.~\ref{sec-envol} and \ref{sec-generalizations}, we showed explicitly that the entropy is submaximal if the universe is spatially curved, of if the free energy is burned up at a time $t_{\rm burn}\ll t_{\rm edge}$ (because there is less free energy in the causal diamond at this time).  In the interest of brevity, we will not consider any of these possibilities further.  Thus, we will only consider examples of processes that are already optimal for entropy production in two respects: they take place during the era $t_\Lambda$, and in a flat FRW universe.

In the first two subsections, we will consider entropy produced as classical radiation, which we define as radiation with a large occupation number $\eta\gg 1$.  In the last two subsections, we will consider quantum radiation, which has $\eta\sim 1$.  In all cases, our only task is to determine the number of quanta, $N$, the occupation number, $\eta$, and the volume fraction, $f_V$.  From Eqs.~(\ref{eq-fdc}) -- (\ref{eq-neta}), and using Sterling's formula, the entropy is
 \begin{eqnarray} 
\Delta S & = & \frac{N}{\eta f_V}\left[-(1-f_V)\log(1-f_V)-f_V\log f_V\right]\nonumber \\
  & + & N\left[-(1-\eta^{-1})\log(1-\eta^{-1})-\eta^{-1}\log \eta^{-1}\right]~.
\label{eq-sneta}
\end{eqnarray} 

Our examples will be given in increasing order of entropy production; note that they all produce less entropy than the maximum possible value\footnote{The value $10^{90}$ is lower than the naive answer, $\Lambda_o^{-3/4}\sim10^{92}$. This is because here we have kept track of all order one factors. We computed the Hubble parameter at the time of maximum comoving volume, $t_{\rm edge}= 0.23 t_\Lambda$, and used the Friedmann equation, $H^2=8\pi \rho/3$, to find the energy density in matter at this time. The quoted value of the entropy is that of blackbody radiation of the same energy density and volume. However, throughout the rest of the section we will neglect order one factors.} 
\begin{equation}
\label{eq-maxent2}
\Delta S_{\rm max}\sim \Lambda_o^{-3/4} \sim 10^{90}~.
\end{equation}

\subsection{Classical radiation from a single source}
\label{sec-c1}

Consider the emission of classical radiation of wavelength $\lambda$, at a time $t_{\rm burn}\sim t_\Lambda$, lasting for a duration $\delta t\ll t_{\rm burn}$.  The energy per emitted quantum is $1/\lambda$.  Therefore, the total number of quanta produced by this process is $N\sim m\lambda$, where $m$ is the total energy emitted.  We will assume that the radiation is emitted incoherently in all angular directions, by a single localized source of size small compared to $\delta t$.  We will evaluate the entropy produced by this source at two times: right after the source shuts off, and just before the radiation leaves the causal diamond.

To avoid complications arising from the dependence of the radiation density on the radius from the source, let us not consider the entropy at the instant when the source shuts off, but at a time of order $\delta t$ after emission has terminated.  By this time, the shell of radiation still occupies a volume of order $\delta t^3$, and its density, of order $m/\delta t^3$, varies at most by a factor of order unity within that volume.  At the time $t_\Lambda$ the volume of the universe (more precisely, of the causal diamond) is of order $t_\Lambda^3$, so the volume fraction occupied by the radiation is $f_V\sim \delta t^3/t_\Lambda^3$.  The density of a single quantum is of order $\lambda^{-4}$, so the characteristic occupation number $\eta$ is of order $m\lambda^4/\delta t^3$.  (We assume that $\eta\gg 1$, i.e., that the radiation is classical.)

Since $\eta\gg 1$ and $f_V\ll 1$, Eq.~(\ref{eq-sneta}) simplifies to
\begin{equation}
\Delta S \sim \frac{N}{\eta} \left(\log f_V^{-1} + \log\eta\right)~.
\label{eq-etabigfsmall}
\end{equation}
The total entropy produced, shortly after the source has turned off, is therefore
\begin{equation}
\Delta S_{\rm early}\sim 
%\frac{\delta t^3}{\lambda^3}\left[\log\left(\frac{m\lambda^4}{\delta t^3}\right)-\log\left(\frac{\delta t^3}{t_\Lambda^3}\right)\right]\sim 
\frac{\delta t^3}{\lambda^3} 
\log\left(\frac{m\lambda^4 t_\Lambda^3}{\delta t^6}\right)~.
\label{eq-searly}
\end{equation}

Now let us consider the entropy of the same radiation at a later time.  While the shell of radiation expands and dilutes, the entropy keeps increasing.  The latest time we can consider is just before the radiation leaves the causal diamond, at a time of order $t_\Lambda$ after it was produced; at later times, any additional entropy increase does not contribute to the quantity we are interested in, the entropy produced {\em within\/} the causal diamond.  At this time, the volume occupied by the radiation is of order $t_\Lambda^2\, \delta t$, so the volume fraction is $f_V\sim \delta t/t_\Lambda\ll 1$.  The number of quanta is unchanged, so the characteristic occupation number is correspondingly smaller: $\eta\sim \lambda^4 m/(t_\Lambda^2 \delta t)$.  (We assume that $\eta\gg 1$ still.)  By Eq.~(\ref{eq-etabigfsmall}), the total entropy produced is
\begin{equation}
\Delta S_{\rm late}\sim 
%\frac{t_\Lambda^2\,\delta t }{\lambda^3}\left[\log\left(\frac{m\lambda^4}{t_\Lambda^2\,\delta t }\right)-\log\left(\frac{\delta t}{t_\Lambda}\right)\right]\sim 
\frac{t_\Lambda^2 \delta t}{\lambda^3} 
\log\left(\frac{m\lambda^4}{t_\Lambda \,\delta t^2}\right)~.
\label{eq-slate}
\end{equation}
Let us apply these results to a specific example.

\paragraph{Black hole merger}  Consider the merger of two supermassive black holes, of total mass $m_b  \sim 10^7 M_\odot$, or $m_b\sim 10^{45}$ in Planck units.  An order-one fraction of this mass, $m\sim m_b$, is converted into gravitational radiation in this process.  We will consider the final stage of the merger, when the two black holes combine into one, and most of the energy is emitted.  The time scale for this process is not much larger than the Schwarzschild radius, $\delta t\sim m_b \sim 100$ sec, and gravitational radiation is emitted in wavelengths not much larger than the Schwarzschild radius, $\lambda\sim m_b \sim 10^7$ km.  Thus, $\delta t/\lambda$ is of order unity, and  from Eq.~(\ref{eq-searly}), we find
\begin{equation}
\Delta S_{\rm early}\sim \log (t_\Lambda^3/m_b)\sim 10^3~,
\end{equation}
Physically, the entropy produced in the merger is miniscule, because although many quanta are produced, they overlap enormously.  The contribution from the large occupation number, $\eta\sim m_b^2\sim 10^{90}$, is log-suppressed because overlapping quanta are indistinguishable.  
%(More, but still not very much entropy is contained in the gravitational radiation produced during a long inspiral process, which we will not consider here.)  
By the time this radiation leaves the horizon, Eq.~(\ref{eq-slate}) yields
\begin{equation}
\Delta S_{\rm late}\sim \frac{t_\Lambda^2}{m_b^2} 
\log\left(\frac{m_b^3}{t_\Lambda}\right)\sim 10^{34}~.
\end{equation}
This entropy is much larger, but still less than the entropy produced by human burning of fossil fuels in the same time frame, about one minute. 

\subsection{Classical radiation from multiple sources}
\label{sec-c2}

Even with many sources, classical radiation is not a very effective way of producing entropy.  We will consider only the case where the entire amount of mass in the causal diamond, $M$, is converted into classical radiation.  (With less mass, the entropy would be even smaller, all other parameters being equal.)  As before, we will assume that the entropy is produced during the era $t_{\rm burn} \sim t_\Lambda$, when $M\sim t_\Lambda $.

Consider $q=M/m$ classical radiators, distributed approximately homogeneously across the universe at the time $t_\Lambda$.  Each radiator emits its entire mass $m$ over a time $\delta t$.  The total number of quanta is $N\sim M \lambda$, and $M\sim t_\Lambda$.   We will only compute the ``late'' entropy, just before all the radiation leaves the horizon.   Since we ignore order one factors, it will not be important that all radiators turn on and off simultaneously, as long as they all operate at some point during the era $t_\Lambda$.  Indeed, we will find that the entropy is independent of both $m$ and $\delta t$.

Naively, the total volume occupied by the multiple shells of radiation is the number of radiators, $q$, times the volume per shell of radiation, $t_\Lambda^2\, \delta t$, which would be $t_\Lambda^3 (\delta t/m)$.  The factor $(\delta t/m)$ is greater than unity (since we assumed that $\delta t$ is greater than the size of the radiator, which in turn must be greater than its mass for gravitational stability), so this volume is much larger than the volume over which the radiation from the whole system of radiators can have spread by causality, $t_\Lambda^3$.   It follows that the $q$ shells overlap by a large factor $(\delta t/m)$.  They cannot be treated as separate; rather, the entropic radiation fills the entire causal diamond homogeneously.  Therefore the volume fraction $f_V\sim 1$, and the occupation number is $\eta\sim \lambda^4/t_\Lambda^2$.  Assuming that the radiation is classical ($\eta\gg 1$), Eq.~(\ref{eq-sneta}) simplifies to
\begin{equation}
\Delta S \sim \frac{N}{\eta} \log\eta~,
\label{eq-etabigf1}
\end{equation}
so the entropy produced is
\begin{equation}
\Delta S_{\rm classical}\sim \frac{t_\Lambda^3}{\lambda^3}\log\left(\frac{\lambda^4}{t_\Lambda^2}\right)~.
\label{eq-sclassical}
\end{equation}

This analysis is valid only if $\eta\gg 1$, which implies
\begin{equation}
\lambda\gg t_\Lambda^{1/2}~.
\end{equation}
Notice that $\Delta S$ is largest at the edge of this regime, when $\eta\to 1$ and $\Delta S\sim t_\Lambda^{3/2}$.  In the opposite limit, when the wavelength is as large as the universe, $\Delta S\sim \log (t_\Lambda^2)$.  Very classical radiation ($\eta\gg 1$) is does not have much entropy.

Let us again apply this result to a specific example: the mergers of supermassive black holes in our universe, as may have happened during hierarchical structure formation, during the era $t_\Lambda$.\footnote{Maor {\em et al.}~\cite{MaoKep08} consider this and other radiative processes but treat overlapping radiation quanta as distinguishable particles, estimating the entropy incorrectly as $S\sim N$.
%\sim M\lambda\sim t_\Lambda \lambda$  
Comparison with Eq.~(\ref{eq-etabigf1}) reveals this to be an overestimate by a factor of $\eta/\log\eta$, a factor of $10^{56}$ in this example.}  For simplicity, we will imagine that all the mass in the universe is in black holes of mass $m_b\sim 10^7 M_\odot$; that all black holes merged pairwise during this era; and that the entire mass of the pair is converted into gravitational radiation.  These assumptions will cause us to overestimate the entropy produced in this process somewhat.   The wavelength is of order the Schwarzschild radius, $\lambda\sim m_b$, and we find from Eq.~(\ref{eq-sclassical})
\begin{equation}
\Delta S_{\rm mergers}\sim 10^{50}~.
\end{equation}
Although this process converts a significant fraction of the mass in the visible universe into entropic radiation, the resulting entropy is negligible compared to many other astrophysical processes.  [The sun alone produces more entropy in a single day; see Eq.~(\ref{eq-ssun}) below.]  This illustrates that very classical radiation ($\eta\sim 10^{57}$ in this example) does not contain much entropy, even if it contains enormous energy.

\subsection{Quantum radiation from a single source}
\label{sec-q1}

Now, consider the emission of quantum radiation by a single source.  By ``quantum'' we mean that the typical occupation number is not very large, $\eta\sim 1$.  (Recall that by definition, $\eta\geq 1$.)  Blackbodies and greybodies are examples of emitters of this type.  In particular, galactic dust clouds fall into this category; as we review in Sec.~\ref{sec-compare}, they dominate the entropy production in our universe, and thus they will play a particularly important role in this paper.

Except for $\eta$, our assumptions will be similar to those of the previous subsections.  We assume that the radiation is emitted roughly isotropically, in the cosmological era $t_{\rm burn} \sim t_\Lambda$, for a duration $\delta t$ 
 into quanta of wavelength of order $\lambda$, by a source small compared to $\delta t$.  The total number of quanta is $N\sim m\lambda$, where $m$ is the total energy emitted.  The total volume available to the radiation is $t_\Lambda^3$. 

The differences to the classical case, Sec.~\ref{sec-c1}, are twofold.  We set $\eta\sim 1$ in Eq.~(\ref{eq-sneta}), corresponding to the assumption that $m\lambda^4\leq \delta t^3$.  Moreover, since quanta overlap negligibly, the volume occupied by quanta is $N\lambda^3\sim m\lambda^4$.  Therefore, the volume fraction is $f_V= m\lambda^4/t_\Lambda^3$.   None of these quantities change while the shell of radiation expands, since the quanta are already dilute to start with; thus, the entropy will be the same right after emission is completed and at the time when the radiation leaves the causal patch.\footnote{If $\delta t^3\ll m\lambda^4\ll t_\Lambda^2 \delta t$, then the radiation is classical ($\eta\gg 1$) at early times but becomes quantum ($\eta\sim 1$) at late times due to the expansion of the shell.  In this intermediate case, Eq.~(\ref{eq-searly}) applies at early times and Eq.~(\ref{eq-quantum1}) at late times.}  In the regime $\eta\sim 1$, $f_V\ll 1$, Eq.~(\ref{eq-sneta}) simplifies to
\begin{equation}
\Delta S\sim N\log f_V^{-1}~,
\label{eq-eta1fsmall}
\end{equation} 
so the entropy produced is
\begin{equation}
\Delta S_{\rm quantum,1}\sim m\lambda \log\left(\frac{\delta t^3}{m\lambda^4}\right)~.
\label{eq-quantum1}
\end{equation}
Let us consider a few specific examples of this type of process occurring in our own universe; in each case, we will consider only a single source of radiation and treat the rest of space as empty.

\paragraph{The sun}  The sun emits at $\lambda\sim 500$ nm, for a time $\delta t\sim 10$ Gyr.  The total energy emitted is $m\approx L_\odot \delta t \approx 7\times 10^{-4} M_\odot$, where $L_\odot\approx 4\times 10^{26} $ W is the solar luminosity.  The total entropy emitted by the sun over its lifetime is 
\begin{equation}
\Delta S_\odot \sim 1.5\times 10^{65}~.
\label{eq-ssun}
\end{equation} 
(This number may seem slightly high, since the entropy is often approximated by the number of photons emitted, $N$.  The log of the volume fraction, which we have included, contributes an extra factor of about 80.)

\paragraph{The galaxy as an optical source}  From afar, our galaxy can be treated as a localized source of visible light---the sum of the emissions from its individual stars.  For a rough estimate, we may use the same parameters as for the sun (previous paragraph), except that the total luminosity (and thus, $m$) is larger by a factor of $10^{11}$. (The Milky Way has about $3\times 10^{11}$ stars, but we shall use a lower number to account for the large abundance of low mass stars that are colder and less luminous than the sun.)  We obtain
\begin{equation}
\Delta S_{\rm gal,opt} \sim 10^{76}~.
\label{eq-sgalopt}
\end{equation} 

\paragraph{The galaxy as an infrared source}  Our galaxy also emits in the infrared, because a significant fraction (neither very close to 1 nor much smaller than 1) of stellar photons are absorbed by galactic dust and re-emitted in the infrared, at $\lambda\approx 200\,\mu$m.  The cosmic background radiation contains about equal power in the optical and in the infrared, so we will use the same total luminosity (and energy, $m$) as for the optical light in the previous paragraph, and $\delta t\approx 10$ Gyr.  We obtain
\begin{equation}
\Delta S_{\rm gal,IR} \sim 2\times 10^{78}
\label{eq-sgalIR}
\end{equation} 
for the entropy produced by the infrared emission from our galaxy.

\subsection{Quantum radiation from multiple sources}
\label{sec-q2}

Now, consider $q$ sources of quantum radiation of wavelength $\lambda$, which we will assume to be distributed approximately homogeneously throughout the universe.    As before, we assume that each source radiates away its entire mass, $m$, approximately isotropically, during the era $t_\Lambda$.  (The duration $\delta t$ of the activity of each source will be irrelevant, except in that we assume the times when each source switches on are roughly evenly distributed over the era $t_\Lambda$.)  We will assume that the entire mass $M\sim t_\Lambda$ of matter inside the causal diamond is in such sources, $qm=M$.  The total number of quanta is $N\sim M\lambda$.

For the entropy radiation to be quantum, $\eta\sim 1$, it is not sufficient that in the absence of any background radiation from other sources, each source would produce radiation with $\eta=1$.   With all sources taken into account, the entropic radiation may have large occupation number, $\eta\gg 1$, by the time most of the quanta have been produced.   There are two regimes, controlled by the parameter 
\begin{equation}
\zeta\equiv \frac{N\lambda^3}{t_\Lambda^3}\sim \frac{\lambda^4}{t_\Lambda^2}~.
\label{eq-zeta}
\end{equation}
If $\zeta\gg 1$, then the radiation becomes dense ($f_V\sim 1$) at the time $t_\Lambda/\zeta$.  At this point, two things can happen: If the sources absorb radiation efficiently, they will equilibrate with one another.  Depending on their dynamical details they will then either heat up, or stop producing any additional quanta.  In this case, we should have described the process as occurring not at $t_\Lambda$ but during the much earlier era $t_\Lambda/\zeta$.  We will not consider this case further.  The other possibility arises if the sources do not absorb radiation efficiently.  In this case, they will keep adding quanta to the dense background, which increases the occupation number to $\eta\sim \zeta$ by the time all sources have shut down.  In this case the final entropy is given by Eq.~(\ref{eq-sclassical}).

If $\zeta\ll 1$, then the radiation remains dilute, $\eta\sim 1$, after all sources have shut down, and the volume fraction is given by $f_V\sim \zeta$.   By Eq.~(\ref{eq-eta1fsmall}), the final entropy is
\begin{equation}
\Delta S_{\rm quantum}\sim t_\Lambda \lambda\log \left(\frac{t_\Lambda^2}{\lambda^4}\right)~.
\label{eq-quantum}
\end{equation}
 
As a more concrete example, consider $q\sim t_\Lambda/m$ spherical blackbodies of radius $R$ distributed at constant density throughout space.  We assume that each blackbody emits its entire rest mass $m$, at constant power, over a time of order $t_\Lambda$.  The temperature $T\sim 1/\lambda$ is determined by
\begin{equation}
\frac{\pi^2}{60} (4\pi R^2) T^4=\frac{m}{t_\Lambda}~.
\label{eq-power}
\end{equation}

We will consider two sets of parameter values.  First, let us set $m\sim 10^{12} M_\odot$ and $R\sim 100$ kpc; from Eq.~(\ref{eq-power}) we find $T\sim 15$ K.  This corresponds to a hypothetical universe filled with homogeneously distributed ``dark matter halos'' of a size comparable to that surrounding the Milky Way.  We assume that these halos emit their entire mass as a blackbody.  In this case, we find $\eta \sim \zeta\sim 10^5$, so the resulting radiation is (marginally) no longer quantum.  From Eq.~(\ref{eq-sclassical}), we find
\begin{equation}
\Delta S_{\rm ``halos"}\sim 10^{88}
\end{equation}
for this process.

For our second example,  let us set $m\sim M_\odot$ and $R\sim R_\odot$; from Eq.~(\ref{eq-power}) we find $T\sim 3\times 10^4$ K.   This corresponds to a (hypothetical) universe filled with homogeneously distributed ``suns'' (which, unlike our sun, burn up their entire rest mass; hence the temperature is higher).  From Eq.~(\ref{eq-zeta}), we find $f_V\sim \zeta\sim 10^{-8}$, so the radiation is quantum and somewhat dilute.  From Eq.~(\ref{eq-quantum}), we find\footnote{It would appear that this example saturates Eq.~(\ref{eq-maxent2}), the maximum entropy that can be produced at all.   Recall, however, that we neglected factors of order one in our examples, whereas Eq.~(\ref{eq-maxent2}) was exact.  With order one factors included, we would find $\Delta S_{\rm ``stars"}\sim 10^{88}$ in this last example.}
\begin{equation}
\Delta S_{\rm ``stars"}\sim 10^{90}~.
\end{equation}

\section{Comparison with observation}\label{sec-compare}

The assumption that observers are correlated to entropy production in the causal diamond led us to predict that observers will find themselves in a flat FRW universe around the time $t_\Lambda$, and that they will see a background of relativistic quanta which were also produced during the era $t_\Lambda$, are nearly space filling, nearly isotropic, and have characteristic wavelength of order $\Lambda^{-1/4}$.  These predictions are obtained independently of the nature of the observers, or the cosmology, or the low-energy particle physics of the vacuum they live in.  In particular, however, they should apply to our own observed universe.  In this section, we will examine whether they are successful.

\begin{enumerate}

\item {\em Prediction}: Most of the entropy production in the causal diamond should take place around the time $t_\mathrm{burn} \sim \Lambda^{-1/2}$; more precisely, the prediction is $t_{\rm burn}\approx t_{\rm edge} \approx .23 t_\Lambda$, which is about $3.5$ Gyr for our value of the cosmological constant, or $10^{60.3}$ in Planck units.\\[1ex]
{\em Observation:}  In the late 1990's the FIRAS and DIBRE instruments on the COBE satellite observed the Cosmic Infrared Background (CIB) radiation\cite{CIB-cobe} in the neighborhood of 100--200 $\mu$m. As shown in Ref.~\cite{BouHar07}, the CIB represents the dominant portion of the entropy produced in our universe within the causal diamond.  (The CMB carries more entropy but is produced much earlier and thus overwhelmingly outside the causal diamond.) The source of the observed CIB is understood to be dominated by starlight from distant galaxies that is attenuated to the infrared by interstellar dust.  Thus, the CIB was emitted predominantly during the era of peak star formation at redshift of 1 to 2, so
\begin{equation}
t_\mathrm{burn}\sim 3.5 \mbox{ - } 6 \mbox{ Gyr}~, 
\end{equation}
depending on the choice of star formation model, or $10^{60.3}$ to $10^{60.5}$ in Planck units.  This is in excellent agreement with the above prediction.

\item  {\em Prediction:} The characteristic wavelength of the entropic radiation is $\lambda\sim \Lambda^{-1/4}$.  With the observed value of $\Lambda$ and order-one factors included, this prediction becomes $\lambda\approx 100\,\mu$m $\approx 10^{30.8}$.\\[1ex]
{\em Observation:} The CIB is not monochromatic, so we need to extract a ``characteristic wavelength'' by an appropriate averaging procedure.  The energy density of the CIB per logarithmic wavelength interval,  $\tilde\rho\equiv\frac{d\rho}{d \log \lambda}$, as measured by the Spitzer Space Telescope~\cite{spitzer}, is shown in Fig.~\ref{fig-spectrum}. %
\begin{figure}
\begin{center}
\includegraphics{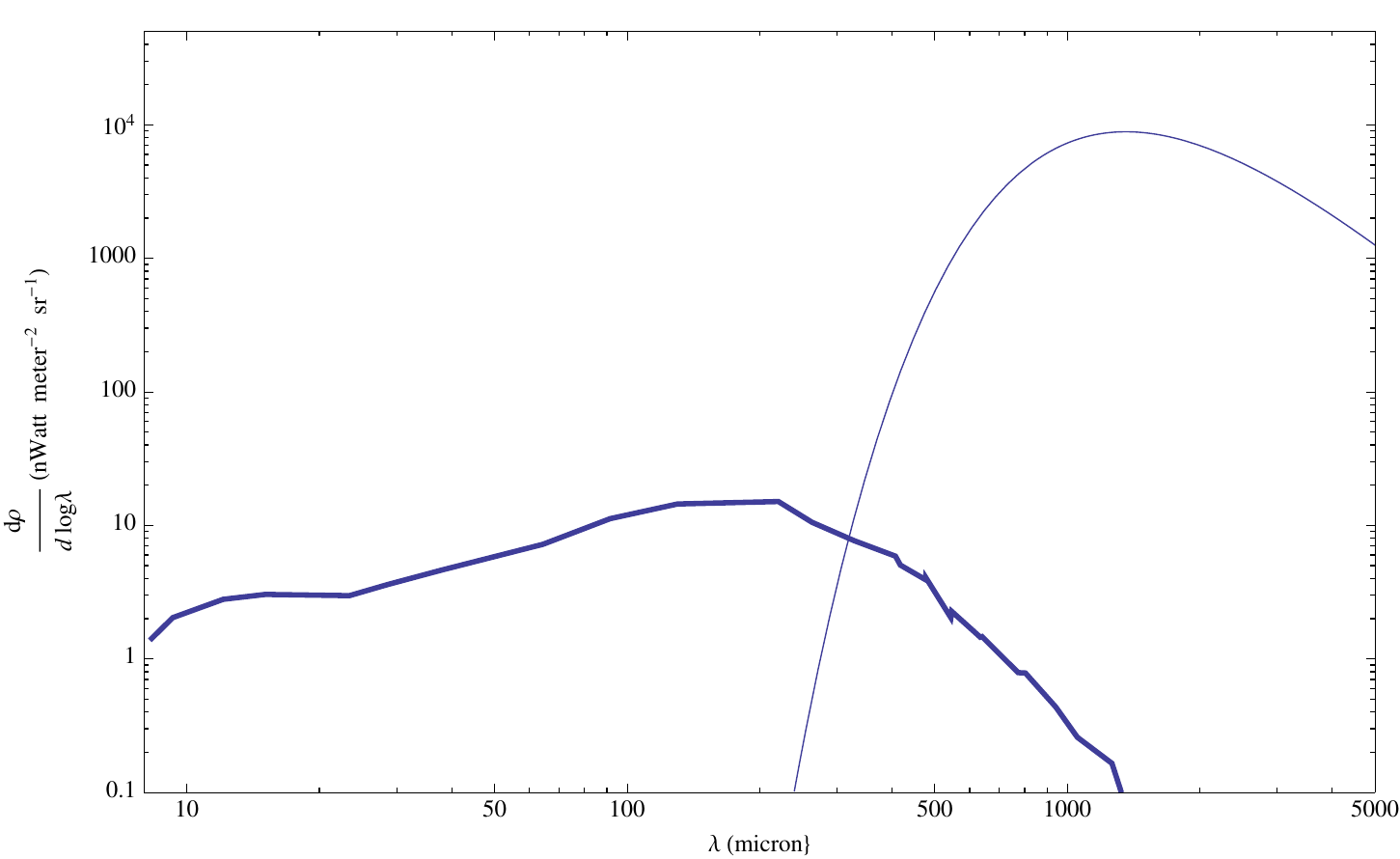}
\caption{The spectrum of the Cosmic infrared background (CIB) taken from~\cite{CIB-cobe,spitzer}. The 2.7~Kelvin microwave background is shown for comparison.}
\label{fig-spectrum}
\end{center}
\end{figure}
In accordance with the entropic principle we will extract the characterstic wavelength of the CIB as weighted by entropy\footnote{Weighting by energy density gives similar results up to order 1 factors.}.  The differential entropy density in the dilute limit is~\cite{planck,lewis}
%\begin{equation}
%s_\lambda=
%\frac{8\pi}{\lambda^4}\left[
%\left(1+\frac{\lambda^5 u_\lambda}{16\pi^2}\right)
%\log\left(1+\frac{\lambda^5 u_\lambda}{16\pi^2}\right)
%-\frac{\lambda^5 u_\lambda}{16\pi^2}
%\log\frac{\lambda^5 u_\lambda}{16\pi^2}\right]
%\end{equation}
\begin{equation}
\frac{ds}{d\log\lambda}=
   \frac{\lambda}{2\pi}\tilde \rho\,
 \log\left(\frac{16\pi^2}{\lambda^4 \tilde \rho}\right)
\end{equation}
The characteristic wavelength of the CIB as weighted by entropy density today is 
\begin{equation}
\langle \lambda \rangle_\mathrm{CIB} =  \frac{1}{\Delta s} 
\int d\log\lambda\,  \lambda  \frac{ds}{d\log\lambda} \approx 290 \mu \mathrm{m} \approx 10^{31.3}~,
\end{equation}
where $\Delta s = \int \frac{ds}{d\log\lambda} d\log\lambda$ is the total entropy density.  This wavelength is in remarkable agreement with the above prediction.
%It should be noted that during the peak of entropy production interstellar dust was slightly hotter because the total stellar luminosity was higher at $z\sim1$-2 by up to an order of magnitude. However, the temperature of dust depends weakly on the flux of the inter stellar radiation field (a power of 1/4 for purely thermal dust, and $\sim1/6$ for a more realistic dust model~\cite{draine}). We thus conclude that the characteristic wavelength was shorter than 290$\mu$m only by a factor of order one. 

\item {\em Prediction:} The entropic radiation fills a significant fraction of the available volume, $f_V\sim 1$. \\[1ex]
{\em Observation:}
The volume fraction taken up by the CIB may be estimated as the ratio of today's CIB flux and the flux of a space filling black body with a similar characteristic wavelength (which in our case is a black body with a temperature of $\sim 18$ K)
\begin{equation}
f_{V} \sim \frac{\rho_{CIB}}{\frac{\pi^2}{15} (18\,\mathrm{K})^4}\sim 10^{-4.3}\,.
\label{eq-fvobserved}
\end{equation}
In evaluating the agreement between prediction and observation, we should keep in mind that $f_V$ has an {\em a priori\/} range of 183 orders of magnitude, so the prediction is relatively precise.  Still, the discrepancy in the case of $f_V$  is larger than for the other five parameters.  We will discusse this issue further in Sec.~\ref{sec-discussion}, where we show that the width of the probability distribution may well be large enough to accommodate Eq.~(\ref{eq-fvobserved}) as a typical value for $f_V$.

\item {\em Prediction:} The entropic radiation is not classical, i.e., the typical occupation number $\eta$ is of order unity\\[1ex]
{\em Observation:}  The CIB is diffuse and is known to originate from nearly thermal dust grains. The emission of these dust grains is far from the classical limit.  
The only way $\eta$ could become much greater than one is if the radiation from many sources overlapped.  If this was then case, then the observed CIB flux per solid angle occupied by galaxies, $\Phi_g$, would exceed the flux of blackbody radiation of the same characteristic wavelength, $\Phi_{\rm BB}$.  But $\Phi_g/\Phi_{\rm BB}$ is observed to be of order $10^{-3}$. Thus the occupation number of CIB photons is low, $\eta\sim 1$, in agreement with the prediction.

\item {\em Prediction:} The universe is spatially flat.\\[1ex]
{\em Observation:} The universe is indeed observed to be spatially flat ($\Omega_k\lesssim 10^{-2}$), as predicted.  (Indeed, a much smaller degree of flatness, $1-\Omega_k\sim O(1)$, would have provided adequate agreement with this prediction.)

\item  {\em Prediction:} The observed entropic quanta are relativistic. \\[1ex]
{\em Observation:} The observed quanta resulting from the dominant entropy production process in our universe are photons and thus relativistic, as predicted.

\end{enumerate}

Above, we have listed the values taken by the scanning parameters in our universe {\em today}.  The entropic principle predicts that most observers live near the time of maximum entropy production.  While this prediction is also quite successful, we do find ourselves living slightly later.  We could instead have listed the values the scanning parameters took at the time of maximum entropy production, roughly at redshift $z\sim1-2$.   However, they are nearly the same.  The volume fraction of existing radiation is invariant under cosmological redshift.  Thus, $f_V$ at redshift $2$ would be only slightly smaller, corresponding to the radiation that was produced after redshift $2$.  The occupation number was $\eta\sim 1$ at all times.  The wavelength $\lambda$ scales as $(1+z)^{-1}$ and so was of order $100\ \mu$m at redshift $2$, in even better agreement with our prediction.

\section{Discussion}\label{sec-discussion}

\paragraph{Does the entropic principle predict that entropy must be produced by galactic dust?} No. While it is remarkable how many successful, quantitative predictions follow from the single assumption of weighting by entropy production, it is important to be clear about what we did {\em not\/} predict.  By the entropic principle, most observers should find themselves in an environment that comes close to maximizing entropy production.   This led us to investigate how entropy production may be optimized in a Universe with a given cosmological constant.  We kept the discussion as general as possible, making no assumptions about the specific dynamics leading to entropy production.  Rather, we asked only questions that are well-defined in arbitrary vacua of the landscape, such as when and at what wavelength the maximum entropy can be produced.  

This allowed us to make predictions for six observable quantities, but not for the mechanism by which the entropy is produced.  We argued only that observers in any vacuum should find that there is {\em some\/} mechanism that achieves parameter values close to those derived in Sec.~\ref{sec-main}, and which thus comes close to optimizing entropy production.   In our universe, this mechanism happens to be the re-radiation of starlight by dust, but this was not one of our predictions.  It could not possibly have been one, since we decided from the start to apply the entropic principle to the landscape as a whole, which meant that such vacuum-specific concepts as ``dust'' and ``stars'' had no place in our analysis.

\paragraph{The entropic principle predicts a flat universe.  So who needs inflation?}  This question is really the same as the previous one.  The entropic principle tells us that a flat universe is more conducive to complex phenomena than a curved one.  It does not tell us how a flat universe is dynamically achieved.  This is similar to how a more traditional anthropic criterion, the requirement for galaxies, implies an upper bound on curvature.  In either case, the argument works only in the context of a landscape that gives rise to spatially flat pocket universes.

Without a specific mechanism, the only way to obtain flatness would be by sheer chance.  This is exponentially unlikely; the small prior probability for flatness would overwhelm the entropic weighting favoring flatness.   Thus, a dynamical mechanism, such as inflation, is needed to explain how some fraction of vacua in the string landscape can become spatially flat.

\paragraph{Five out of our six predictions agree perfectly with the data.  How unhappy should we be that one parameter is four orders of magnitude off?}  Let us ignore, for a moment, the successful predictions of spatial flatness and of the relativistic nature of entropic radiation, which are of a more qualitative nature.  Three of the remaining four predictions (the time of entropy production, the typical occupation number, and the wavelength) agree with the respective observed values to within less than an order of magnitude.   The fraction of volume occupied by the infrared background radiation in our universe, however, is observed to be $f_V\approx 10^{-4.3}$; the predicted value is about one.   

A discrepancy of four orders of magnitude looks bad until we remember that we have identified only the most likely values of parameters, but not the width of the peak around them.  Weighting by entropy production comes on top of an underlying statistical vacuum distribution of the multiverse.  We have assumed that this underlying distribution does not grow so strongly towards some other parameter values as to overwhelm the weight coming from entropy production.  However, the underlying distribution can sharpen or widen the peak that would naively be obtained from a flat prior over the logarithm of our parameters.  In other contexts~\cite{BouHal09}, deviations of order $10^4$ in certain parameter combinations were found to be natural for a plausible range of prior distributions.  Thus, the only concern is that our predictions may be so uncertain as to be vacuous.  However, even a fairly wide peak of four orders of magnitude in $f_V$ (under which the observed value would be highly typical, i.e., in the $1\sigma$ region) would still constitute a rather restrictive prediction, since $f_V$ can vary {\em a priori\/} over $183$ orders of magnitude in vacua with $\Lambda=\Lambda_o$.  In other words, the volume fraction is still remarkably close to a special value.

In any case, it is best to regard our result as a joint prediction of (at least) the four parameters $(f_V,\eta,t_{\rm burn},\lambda)$.  The combined discrepancy is only about an order of magnitude per parameter, in a four-dimensional logarithmic parameter space of volume of roughly $63\times61\times183\times122\sim 10^7$.  Given the simplicity of the causal entropic principle, we regard this as a remarkable success.

\paragraph{Doesn't this result mean that we are atypical at the level $10^{-5}$?} 
We have identified the main inefficiency in how our universe produces entropy: the volume fraction occupied by entropic radiation is only $.5\times 10^{-4}$.  We have just argued that the probability distribution can be so wide as to make this result quite typical.  But how can this be?  Aren't observers $10^4$ times more likely to find themselves in a universe with volume fraction $f_V=1/2$?  This concern is naive, because prior landscape distributions must be taken into account.  To see this in an explicit example, let us consider what would happen if we try to design a vacuum ``like ours'' in which the volume fraction occupied by entropic radiation is larger.   

We would like to hold fixed all other parameters, which are already (nearly) optimal: the time of dominant entropy production, $t_{\rm burn} \sim 10$ Gyr, the diluteness of the radiation, $\eta\sim 1$, and the characteristic wavelength, $\lambda\sim 100\,\mu$m, the flatness of the universe, and the relativistic nature of the entropic radiation.  Then the only way to increase the volume fraction is to burn up a greater fraction of the rest mass of the universe.

In an optimal Universe all of the rest mass  is converted into entropic radiation. Our Universe is not as energy efficient. Only about a sixth of matter is baryonic. Only about 10\% of baryons cooled and fell into stars. Stars burn only 10\% of their baryons. Restricting to baryons that participate in stellar burning, only 0.007 of their mass is converted into radiation when four hydrogen nuclei fuse into Helium-4. Finally, only an order one fraction of stellar photons are reprocessed by dust into IR photons.  This cumulative inefficiency factor of about $10^{-5}$ essentially accounts for the five order of magnitude discrepancy from optimal entropy production. 

But we cannot improve entropy production merely by making our Universe more energy efficient. For example, consider the free energy fraction in nuclear reactions, 0.007, which contributes the most of all the inefficiencies we have listed. Had this parameter been larger (by as little as a factor of two, if we keep all other parameters fixed), most of the hydrogen in the Universe would have burned into the now-stable $^2\mathrm{He}$~\cite{anthropicBBN}. This would change other parameters away from their optimal values: the time of dominant entropy production would be a fraction of a minute, rather than 10 billion years; and the characteristic wavelength would be much smaller than in our universe.  (Note that the entropic principle gets it right: such a universe would seem unlikely to harbor observers.)

Alternatively, increasing the fraction of baryons in galaxies from 10\% to 100\% (by some variation of particle physics parameters) would seemingly allow us to gain an order of magnitude in entropy production. However, the increased radiation pressure from the additional starlight and infrared radiation would blow the dust out of the galaxies.  This would result in a loss of the factor of about $100$ by which the entropy is increased in our universe when starlight is scattered by dust. 

We conclude that while it may be possible to increase the entropy production by another four orders of magnitude, doing so requires careful tuning of many landscape parameters, and thus suffers from statistical suppression.  This illustrates why the width of the distribution may well be large enough to accommodate the observed deviation from optimum entropy production.

\paragraph{Does the entropic principle make other predictions?}
We have taken a very general approach, referring only to concepts that can be defined in all semi-classical landscape vacua.   Further predictions may be made at the expense of restricting to more specific classes of vacua.   Here we will mention one possible direction for progress: the temperature of the CMB and properties of dust.  In Figure~\ref{fig-spectrum} we showed the spectral peak  due to the infrared emission by dust that is heated by starlight. 
This peak is less than one order of magnitude from the black body peak of the cosmic microwave background; indeed, the two spectra partially overlap.  This is a surprising coincidence, because the temperature of dust is not related dynamically to the CMB temperature.

The wavelength of the CIB is determined by the strength of the interstellar radiation field in galaxies, as well as by the absorption and emission properties of the dust grains that scatter the starlight. Very roughly, if dust were a perfect black body (we will consider deviations from this below) the temperature of dust $T_\mathrm{dust}$ would be set by requiring that the absorbed and emitted fluxes are equal:
\begin{equation}\label{eq-Tdust}
4\pi a^2 T_\mathrm{dust}^4 \approx \sigma_a (\Phi_\star + \ldots)~,
\end{equation}
where $\Phi_\star$ is the average energy flux of starlight in the galaxy, $\sigma_a$ is the absorption cross section, and $a$ is the radius of a dust grain. 
Assuming that the absorption cross section is geometrical, we get $T_\mathrm{dust}\sim \Phi_\star^{1/4}$.  In Eq.~(\ref{eq-Tdust}), we have omitted sources of radiation whose contribution is subdominant to those of starlight.  In our universe, the 2.7 K background does not play a role in setting the wavelength of the CIB because it is a subdominant contribution. However if the CMB temperature were just an order of magnitude higher, its flux would begin dominating the sum on the right hand side of equation~\ref{eq-Tdust}, $\Phi_\star+T_\mathrm{CMB}^4$.  The sensitivity to the CMB temperature is strong because of the high power with which it enters the equation.  (In a more realistic description the right hand side depends on an even higher power of $T_\mathrm{CMB}$.) In practice, the temperature of dust is set by the largest absorption term. It is thus possible to explain the value of the temperature of the CMB in our Universe by assuming a landscape pressure towards high $T_{CMB}$.  The value we observe is the largest value that does not affect the value of $\lambda$, the wavelength of the infrared background, which is already optimal in our Universe.  In other words, the observed CMB temperature is very close to the critical value at which dust would be heated as much by the CMB as by starlight.  A higher CMB temperature would result in a higher dust temperature and thus in less entropy production.

Further predictions of fundamental parameters may be made because interstellar dust is not a black body, and both emission and absorption rates depend on wavelength and other dimensionful quantities.  The temperature of dust is thus affected by microphysical parameters. For example, a more realistic but simple model of dust as a grey body gives~\cite{andriesse, draine}
\begin{equation}
T_\mathrm{dust}\approx \left[ \omega_p ( T_\star \Phi_\star + T_\mathrm{CMB}^5)\right]^{1/6}
\end{equation}
where $\omega^2_p\sim n/m_N$ is the plasma frequency within a dust grain ($n$ is the number density of atoms in a dust grain, of order $10^{24}$ cm$^{-3}$  and $m_N$ is the nuclear mass). The mild sensitivity to fundamental parameters such as the electron and proton mass may place some restrictions on their values from entropic considerations. Other quantities that may be constrained in this way are the galactic opacity in the IR (which is of order one in our Universe, a coincidence) and the dust-to-gas ratio. Since all these predictions arise in a more restrictive setting than that considered here, we will not consider them further in this paper.

\paragraph{What if we allow $\Lambda$ to scan?}

We have used the cosmological constant as an arbitrary, but fixed, input parameter and obtained predictions for six scanning parameters in terms of $\Lambda$.  This is legitimate; but we can go further and ask what values of $\Lambda$ are most likely to be observed, assuming nothing else.  What happens if we allow all parameters to scan, including $\Lambda$?

What happens depends critically on the underlying probability distribution in the landscape of vacua. This is reliably known only for $\Lambda$ itself: Since $\Lambda=0$ is not a special point from the particle physics point of view, the prior probability distribution is $d\tilde p/d\Lambda\approx \mbox{const}$ for $|\Lambda|\ll 1$ by Taylor expansion.  Combing this with the entropic weighting obtained after integrating out all other parameters, $\Delta S\sim \Lambda^{-3/4}$, and assuming a flat prior on the logarithm of all other parameters, we obtain
\begin{equation}
\frac{dp}{d\Lambda}\propto \Lambda^{1/4}~.
\label{eq-lambda14}
\end{equation}
This would still favor large values of $\Lambda$, suggesting that we are the least complex observers possible.  Our intuition rebels against this conclusion, though we have not found a sharp argument that it conflicts with observation.

However, the conclusion is far from inevitable, since the prior distributions may well be different.  We will focus on a particular parameter, the time $t_{\rm burn}$, when most entropy is produced (or, via the entropic principle, when observers live).   It may well be easier to produce more entropy at late times.  In other words, there may be a landscape force favoring large values of $t_{\rm burn}$ (or more generally, large values of $t_{\rm obs}$, if we were modelling observers not by entropy production but in some other way).  This is simply the statement that it takes a long time to make complex things, especially if they emerge from an unguided process.  Let us express this assumption as
\begin{equation}
\frac{dp}{d t_{\rm burn}}\propto t_{\rm burn}^\alpha~,
\end{equation}
where $\alpha$ is a positive number of order unity.

We have shown in Sec.~\ref{sec-main} that optimizing entropy production sets $t_{\rm burn} \to t_\Lambda$.  Thus, after integrating out $t_{\rm burn}$, the above pressure towards large values of $t_{\rm burn}$ is inherited by $t_\Lambda$,  adding $\alpha$ to the probability distribution over $t_\Lambda\sim |\Lambda|^{-1/2}$.   Thus, Eq.~(\ref{eq-lambda14}) becomes
\begin{equation}
\frac{dp}{d\Lambda}\propto |\Lambda|^{\frac{1}{4}-\frac{\alpha}{2}}~.
\label{eq-lambdaalpha}
\end{equation}
If $\alpha>1/2$, which seems to us far from implausible, the probability distribution favors {\em small\/} values of the cosmological constant.\footnote{The special case $\alpha=1/2$ leads to a constant probability distribution over $\log|\Lambda|$.  The conclusions in this case are similar to the case $\alpha>1/2$, except that our value of $\Lambda$ would be related only to the order of magnitude, not to the actual value, of the smallest $|\Lambda|$ in the landscape.}  In this case, we would find ourselves at the ``discretuum limit''.  The cosmological constant we observe is controlled by the smallest value of $|\Lambda|$ in the whole landscape, which in turn is roughly the inverse of the number of vacua with observers.   

This possibility strikes us as extremely interesting, in that it relates the observed value of $\Lambda$ directly to the size of the string landscape.  It is particularly intriguing that some estimates of the number ${\cal N}$ of metastable vacua have given $\log_{10} {\cal N} \sim 500$~\cite{BP,DenDou04b}.  Note that the number of anthropic vacua may be a very small fraction (perhaps $10^{-O(100)}$ of all vacua.  So we should not conclude that we live in the vacuum with the smallest cosmological constant; we can conclude only that the value we observe is related to this smallest value by some factor that captures the rarity of vacua with particle physics and cosmological parameters that admit complex phenomena such as observers.  From purely mathematical facts (the number 500 is related to the number of distinct topological cycles in certain manifolds), this could explain the origin of a kind of ur-hierarchy in physics, from which all other hierarchies may be derived either by symmetry arguments of anthropic correlations.

\paragraph{Can the entropic principle be improved?}  We hope so, and leave this to future work.  Clearly, entropy production is an imperfect proxy for observers.  It is easy to imagine processes that optimize entropy production but are not accompanied by any kind of complex phenomenon.  Worse, such processes may be the norm.  For example, one would expect that for every vacuum like ours, with complex galaxies, long-lived stars, planets, and organic molecules, there are many vacua (with the same cosmological constant) in which equal or more entropy is produced by the decay of a single neutrino species with mass 3 meV and lifetime 10 Gyr.   These vacua should be more abundant because unlike in our vacuum, the dominant entropy production process would not be predicated on the complex interplay (and concomitant tuning) of a large number of particle physics parameters.  Despite the near-maximal entropy production, this process would be very primitive, with each neutrino producing a couple of photons or other decay products, and no complex structures of any type.

This is not necessarily a problem.  Our predictions were successful anyway, because they did not depend on assuming that entropy production is sufficient for observers.  But the entropic principle does assume something stronger than the innocuous statement that entropy production is necessary:  It posits that the number of observers is {\em proportional\/} to the amount of entropy produced.  This is likely to work best if we average over (i.e., integrate out) most landscape parameters, as we have done here.  In other words, the fact that we considered the whole landscape was actually beneficial.  

In a more restrictive setting, the entropic principle has greater limitations.  For example, if we focussed on a universe like ours except for the value of $\Lambda$, but (unlike Ref.~\cite{BouHar07}) allowed $\Lambda$ to scan over a range large enough to include $t_\Lambda \sim t_{\rm reheat}$, the probability distribution over $\Lambda$ would be dominated by a second peak near $\Lambda\sim t_{\rm reheat}^{-2}$ due to the entropy production at reheating.  The challenge, then, will be to find a proxy for observers that captures the notion of ``complexity'' more reliably, while preserving the simplicity, generality, and phenomenological successes of the entropic principle.   It is possible that some of the criteria developed in the existing literature on complexity will be of use here.
 
\appendix
\section{The causal diamond measure}
\label{sec-measure}

\begin{figure}
\begin{center}
\includegraphics[width=.8\textwidth]{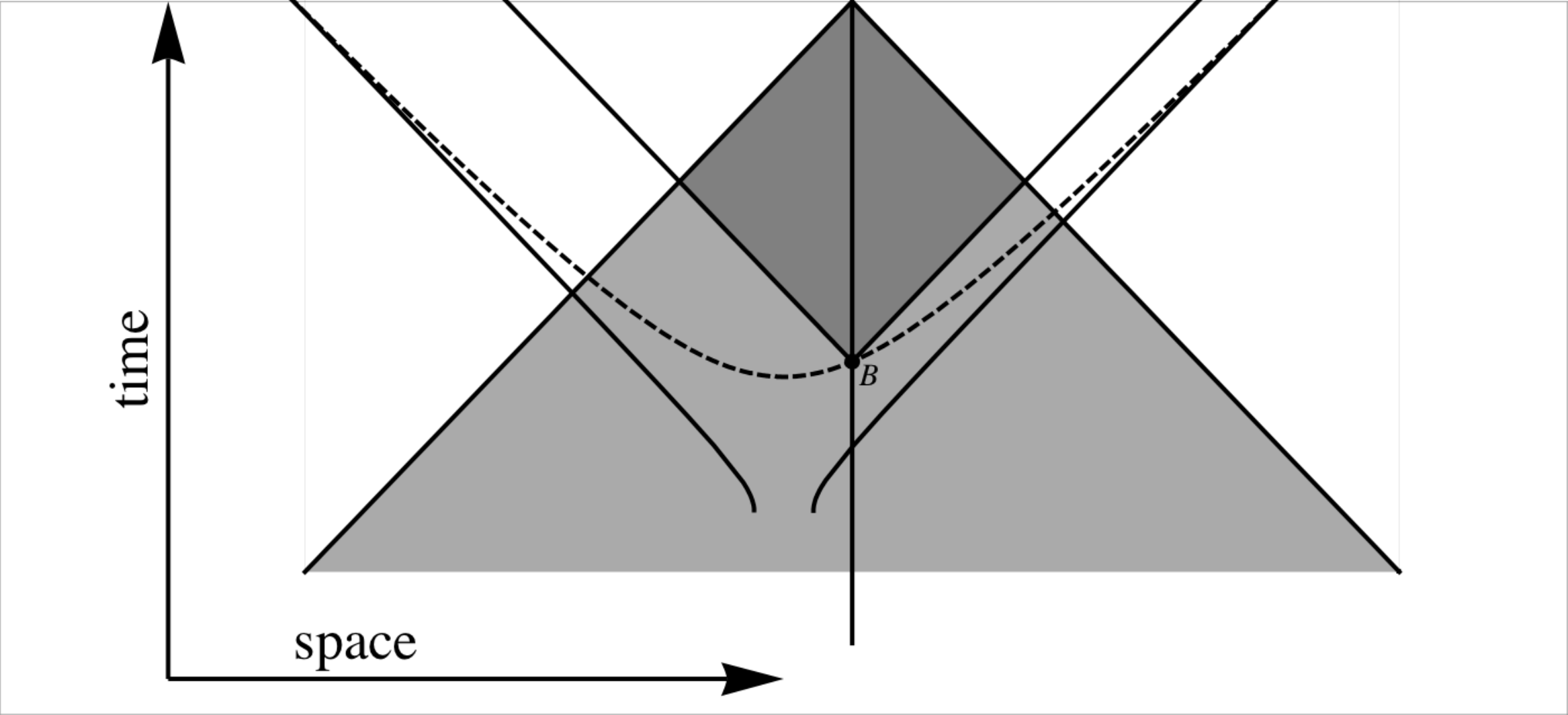}
\caption{The causal patch (shaded triangle) is the past of the
  future endpoint of a geodesic (vertical line) in
  the multiverse.  In this paper we use the causal diamond measure.  The causal diamond (dark shaded) is the intersection of the
  causal patch with the future of the point $B$, where the geodesic
  intersects a surface of reheating (dashed).  A domain wall separating two vacua is also shown (the large wedge that is open at the bottom).}
\label{fig-cd}
\end{center}
\end{figure}
In a universe with at least one metastable de~Sitter vacuum, inflation is eternal.  All events that can happen at all will happen infinitely many times.  To define relative probabilities, these infinities must be regulated.  This is the measure problem. The landscape of string theory, which contains many metastable de~Sitter vacua, has focussed renewed attention on this old problem.  

A number of simple measure proposals have been ruled out by showing that they conflict dramatically with observation.  Some of the most promising surviving proposals have been shown to be exactly equivalent due to global-local dualities, or differ only by relatively small factors~\cite{BouFre08b,Bou09,BouYan09}.  These include the causal diamond and causal patch measures (with certain simple initial conditions, the latter is equivalent to the global light-cone time cutoff), as well as the ``fat geodesic'' measure (which is equivalent to the global scale factor time cutoff in regions where the latter is well-defined).  These measures are generally in good agreement with observation~\cite{BouHar07,BouFre08b,Cline,Albrecht, BouLei09}.

The causal patch measure~\cite{Bou06} restricts attention to the causal past of a timelike geodesic in the eternally inflating multiverse (or equivalently, of the ``endpoint'' of the geodesic on the conformal boundary of spacetime).  In this paper, we use the causal diamond measure~\cite{BouHar07} (see Fig.~\ref{fig-cd}), which is more restrictive.  The causal diamond is the intersection of the causal patch with the causal future of the event where the geodesic intersects a reheating surface.  (The definition has the disadvantage that the term ``reheating surface'' is not as sharply defined as one would like for a general measure.  This limitation can be overcome by using the apparent horizon measure, which restricts to the spacetime region covered by the union of the expanding (in the past direction) portions of the past light-cones of all points on the geodesic.)

If we had used the causal patch measure instead of the causal diamond, we would find that entropy production in our universe is dominated by the CMB.  In Sec.~\ref{sec-discussion} we argued that the decay of a particle is not a form of entropy production associated with truly complex phenomena, since the structures involved consist of only a few bits each and are essentially independent.  Thus it is possible that this problem can be overcome with an improved proxy for observers that captures complexity more reliably.  Another problem is that for $\Lambda<0$, the causal patch favors large spatial curvature and/or a small magnitude of the cosmological constant~\cite{BouLei09}. Indeed, the validity of the causal patch cutoff, or of any related proposals, in non-eternally inflating regions is questionable; this issue may have to await a better understanding of spacelike singularities.

It is worth emphasizing that the choice of measure is independent of how we model observers (see, however, footnote \ref{fn}).  The purpose of the measure is to remove divergences, which arise in eternal inflation independently of how observers are modelled.

\acknowledgments We are grateful to M.~Arvanitaki, M.~Davis, S.~Dimopoulos, S.~Dubovsky, D.~Fink\-beiner, B.~Freivogel, P.~Graham, L.~Hall, S.~Leichenauer, Y.~Nomura, P.~Richards, and U.~Seljak for discussions.  The work of RB was supported by the Berkeley Center for Theoretical Physics, by the National Science Foundation (award number 0855653), by the Institute for the Physics and Mathematics of the Universe, by fqxi grant RFP2-08-06, and by the US Department of Energy under Contract DE-AC02-05CH11231.

\bibliographystyle{utcaps}
\bibliography{all}
\end{document}